\documentstyle[preprint,aps]{revtex}
                    \tighten
                    \begin{document}
                    \draft
                    \title{Transmission and Spectral Aspects of Tight
          Binding Hamiltonians For the Counting Quantum Turing Machine} 
                    \author{Paul Benioff\\
                     Physics Division, Argonne National Laboratory \\
                     Argonne, IL 60439 \\
                     e-mail: pbenioff@anl.gov}
                     \date{\today}

                    \maketitle
                    \begin{abstract}  
          1-D systems with deterministic disorder, such as those with
          quasiperiodic or substitution sequence potential distributions,
          have been much studied.  It was recently shown that a
          generalization of quantum Turing machines (QTMs), in which
          potentials are associated with elementary steps or transitions of
          the computation, generates potential distributions along
          computation paths of states in some basis $B$, that are 
          computable and are thus periodic or have deterministic disorder. 
          These generalized machines (GQTMs) can be used to investigate the
          effect of potentials in causing reflections and reducing the
          completion probability of computations. This paper expands on
          this work by determining the spectral and transmission properties
          of an example GQTM, which enumerates the integers in succession
          as binary strings.  A potential is associated with just one type
          of step.  For many computation paths the potential distributions
          are initial segments of a distribution that is quasiperiodic and
          corresponds to a substitution sequence.  Thus the methods
          developed in the study of 1-D systems can be used to calculate
          the energy band spectra and Landauer Resistance (LR). For
          energies below the barrier height, The LR fluctuates rapidly with
          momentum with minima close to or at band-gap edges.  Also for
          several values of the parameters involved there is good
          transmission over some momentum regions.

          \end{abstract}

          \section{Introduction}
          The discovery of quasicrystals \cite{Sh}, (See \cite{FuO,DiVSt}
          for recent reviews) has stimulated much study of systems with
          deterministic disorder \cite{BoGh1}. These systems are
          neither random nor periodic.  As part of this work there has been
          much interest in the study, either with Kronig-Penney or tight
          binding Hamiltonians, of the spectra and transmission properties
          of 1-D systems whose potential distributions correspond to
          substitution sequences. (These are defined in \cite{defsub}). 
          For the Period Doubling, \cite{BoGh1,Luck,BeBoGh}, Thue-Morse
          \cite{BoGh1,Luck,BeBoGh}, and Fibonacci \cite{Suto} substitution
          sequences, the spectra are singular continuous and are Cantor
          sets.  For the Rudin-Shapiro sequence, the spectral properties
          are unknown \cite{BoGh2}.  Potential distributions corresponding
          to circle sequences \cite{Horn} and the prime number distribution
          \cite{Ryu} have also been studied. 

          Transmission properties of 1-D structures for which the potential
          distributions are initial segments of substitution sequences have
          also been studied \cite{AvBeGl,AvBe1,AvBe2,GaKh,Erdos1}. Of
          special interest here are studies of the Landauer Resistance
          \cite{Landauer,Erdos} for various potential distributions
          and its relation to the band structure of infinite
          periodic systems whose unit cells are initial segments of
          substitution sequences \cite{Roy1,Roy2,RoyBasu}.

          One dimensional structures of these types also appear in a
          generalization of quantum Turing machines introduced elsewhere
          \cite{Benioff3}.  The generalization consists of modifying the
          Hamiltonian description of quantum Turing machines (QTMs) so that
          potentials are associated with some types of steps or transitions
          of the computation as it moves along paths of states in a
          suitable basis.  The resulting distributions of potentials along
          these computation paths are computable and are thus either
          periodic or deterministically disordered.  As a result
          generalized quantum Turing machines (GQTMs) are both quantum
          computers and 1-D structures of the type of much recent interest.

          The presence of these potentials means that GQTMs also serve as a
          test bed to examine the effect of these potentials on the
          transmission and reflection of the computation state along any
          computation path and the completion probability for the
          computation. The importance of this for quantum computation has
          been repeatedly emphasized  by Landauer \cite{Land1}.  He has
          noted that these potentials, which may result from environmental
          influences and errors in construction of actual physical models
          of the computation process \cite{Peres,Zurek}, cause reflections 
	  back along the computation path and may result in exponential decay 
	  or localization of the transmitted state.  

          This emphasis is based on the fact that in much of the work done
          so far on quantum computation, the computation as been assumed to
          proceed smoothly along computation paths with no potentials
          present. This is the case whether quantum computers are used as
          computers 
          \cite{Benioff1,Benioff,Deutsch,Lloyd1,Div,ChLa,Unr,LaMi} or as
          simulators of other quantum systems \cite{Feynman1,Lloyd2}. Work
          on Shor's quantum algorithm \cite{Shor} and on quantum gates and
          quantum error correction \cite{Shor1,Lloyd} is also included.
	  
	  The main purpose of this work is the examination of the effect of
potentials on the spectral and transmission properties of computation
states along computation paths.  The method will be based on other work 
   \cite{Benioff3} in which potentials are 
          associated with some types of steps or transitions
          of the computation as it evolves along paths of states in a
          suitable basis.  The Feynman
          Hamiltonians for these GQTMs can, under certain conditions,
	   be decomposed into a direct sum of tight
          binding Hamiltonians, $H=\sum_{i}H_{i}$, where the $i$ sum is over
          computation paths of states in a basis.  The potential
          distribution, $V_{i}$, in $H_{i}$ depends on the path $i$.  
	  
	  Here the example GQTM, which was introduced in other work 
	  \cite{Benioff3}, will be examined in detail. This
          example, the counting GQTM, which enumerates the integers in
          succession as binary strings, has a potential
          associated with read-1 steps only.  (read-1 steps are defined
          at the end of Section \ref{TBH}.) For many paths the potential
          distribution was seen to be quasicrystalline and to correspond to
          initial segments of a substitution sequence generated from an
          infinite alphabet.   

          Both the spectral and transmission properties of the tight binding 
          Hamiltonians associated with the counting GQTM will be determined. 
	  The energy band spectra will be calculated for associated periodic 
	  systems. These are 1-D crystals whose unit cells are (finite 
	  sections of) computation paths that contain the potential 
   distributions being studied.  Transmission properties will be calculated by
          determining the Landauer Resistance of computation paths for the
          counting GQTM.  This paper should be read in conjunction with 
          \cite{Benioff3} as much of the earlier work will not be repeated.

          In the next section the physical model for QTMs (generalized or
          not) will be presented. This will be followed in Section
          \ref{TBH} by a brief description of the generalization of QTMs
          and the decomposition of Hamiltonians for GQTMs into a sum over
          paths of tight binding Hamiltonians.  Computation paths will be
          defined.   The counting GQTM will be described in Section
          \ref{CGQTM}.  The discussion of the step operator and potential
          distribution will be brief as much of the material is already
          given elsewhere \cite{Benioff3}.  The transfer matrix will be
          defined for potential distributions containing the first $2^{N-
          1}$ potentials with $N$ an arbitrary positive integer.

          The emphasis of Sections \ref{SP} and \ref{LR} is on the spectral
          gaps and band widths for the associated periodic systems and
          Landauer Resistance of systems described by tight binding
          Hamiltonians with $2^{N-1}$ potentials for the counting GQTM. The
          width and separation of the potentials is related to an initial
          segment of a generalized substitution sequence described
          elsewhere \cite{Benioff3,Turban}. The work is exploratory with
          emphasis on calculations to show the dependence of these
          properties on various parameters.  Also as was done by Roy and
          Khan \cite{Roy2}, the relation between the Landauer Resistance of
          a system and the band spectra of the associated periodic system
          will be investigated. 

          It should be noted that the only effect of this generalization is
          the introduction of potentials along computation paths.  There is
          no change in the class of QTMs or in the concept of quantum
          computability. 

          \section{The Physical Model}
          \label{TPM}
          The physical model for one-tape quantum Turing machines consists
          of a two-way infinite one dimensional lattice of systems each of
          which can assume states in a finite dimensional Hilbert space. 
          If the space is two dimensional, the systems are referred to as
          qubits.  This term will be used here even if the dimensionality
          is greater than two.  It is often convenient but not necessary to
          consider the lattice as spin systems, e.g. spin 1/2 systems for
          binary qubits.

          A head which can assume any one of a finite number of orthogonal
          states $\vert l\rangle$ with $l\epsilon L$ moves along the
          lattice interacting with qubits at or next to its location on the
          lattice.  Elementary QTM actions include one or more of (1) head
          motion one lattice site to the right or left, (2) change of the
          state of the qubit scanned or read by the head, (3) change of the
          head state.  What happens depends on the states of the head and
          scanned qubit. 

          Here the system states are all assumed to lie in a separable
          Hilbert space $\cal H$.  Based on the above description a
          particular basis, the computation basis, defined by the set of
          states $\{\vert l,j,S\rangle\}$ and which spans $\cal H$, is
          used.  Here $l,j$ refer to the internal state and lattice
          position of the head.  The qubit lattice computation basis state
          $\vert S\rangle =\otimes_{m=-\infty}^{\infty} \vert S(m)\rangle$
          where $S$ is a function from the integers to the qubit state
          labels (e.g. $\{0,1\}$) such that  $S(m)\neq 0$ for at most a
          finite number of values of $m$.  This condition, the $0$ tail
          state condition, is one of many that can be imposed to keep the
          basis denumerable.  Note that all qubit states, such as $ a\vert
          0\rangle +b\vert 1\rangle$ for arbitrary $a,b$ and sums of
          products of these states are included as appropriate linear
          superpositions of  the computation basis states.

          \section{Tight Binding Hamiltonians from Quantum Turing Machines}
          \label{TBH}

          One begins by associating a bounded linear operator $T$ with each
          GQTM.  $T$ is referred to as a step operator for the GQTM because
          iteration of $T$ (or its adjoint) corresponds to the successive
          steps in the forward (or backward) time direction of the GQTM. An
          infinitesimal time interval is associated with the steps of $T$
          so that it can be used directly to construct a time independent
          Hamiltonian. The specific construction used here is that proposed
          by Feynman \cite{Feynman}:
          \begin{equation}
          H=K(2-T-T^{\dag}) \label{ham}
          \end{equation}
          where $K$ is a constant.  This definition has the advantage that
          if $T=Y$ where $Y$ is the bilateral shift along the lattice, then
          $H$ is the kinetic energy of free head motion on the lattice.  As
          such it is equivalent to the symmetrized discrete version of the
          second derivative, $(-\hbar^{2}/2m)d^{2}/dx^{2}$. 
	  
          For each GQTM the associated step operator $T$ is defined as a
          finite sum over elementary step operators.  That is
          $T=\sum_{l,s}T_{ls}$ where the sum is over all head state labels
          $l\epsilon L$ and all qubit state labels $s\epsilon \cal S$ where
          $\cal S$ is the set of qubit state labels.  For binary qubits
          ${\cal S} =\{0,1\}$.  $T_{l,s}$ corresponds to the action taken by
          the GQTM associated with $T$ when the head in state $\vert
          l\rangle$ sees or reads a qubit in state $\vert s\rangle$.

          The generalization consists in noting that for each $l,s$
          $T_{l,s}=\gamma_{l,s}W_{l,s}$ where $\gamma_{l,s}$ is a positive
          real constant $\leq 1$.  For QTMs without the generalization
          $\gamma_{l,s} =1$ for all $l,s$.  $W_{l,s}$ is defined as a sum
          over all lattice positions of products of projection operators,
          bit and head state change, and head position change operators. 
          The specific definition of $W_{l,s}$ appears elsewhere
          \cite{Benioff3}.

          From now on either the $W_{l,s}$ or the $T_{l,s}$ will be
          referred to as the elementary step operators for the GQTM.  The
          reference wiil be made explicit in contexts where the difference
          between $T_{l,s}$ and $W_{l,s}$ is important.  

          The main condition imposed on $T$ and its adjoint is that they be
          distinct path generating in some basis $B$ which spans $\cal H$. 
          This means that iterations of $T$ or $T^{\dag}$ on any basis
          state in $B$, generate a path of states that, up to
          normalization, are also states in $B$.  Furthermore the paths are
          distinct in that no two paths join, intersect, or branch.  

          Although $B$ can be any basis which spans $\cal H$, (see
          \cite{Benioff} for details), $B$ is required here to be the
          computation basis, defined in Section \ref{TPM}.  A path in 
          $B$ is defined as a set of states in $B$ ordered by iteration of
          $T$ or $T^{\dag}$.  If the state $\vert l_{m},j_{m},S_{m}\rangle
          =\vert m,i\rangle$ is the $mth$ state in some path $i$, then the
          $m+1st$ and $m-1th$ states in the path are given respectively by
          $\vert m+1,i\rangle = W_{l_{m},S_{m}(j_{m})}\vert
          l_{m},j_{m},S_{m}\rangle = \vert l_{m+1},j_{m+1},S_{m+1}\rangle$
          and $\vert m-1,i\rangle = W^{\dag}_{l_{m-1},S_{m-1}(j_{m-
          1})}\vert l_{m},j_{m},S_{m}\rangle = \vert l_{m-1},j_{m-1},S_{m-
          1}\rangle$.   These equations must be modified if either $T$ or
          $T^{\dag}$ annihilate $\vert m,i\rangle$.

          Mathematically the requirement that $T$ be distinct path
          generating in the computation basis, $B$, can be expressed by the
          condition that $T$ is a direct sum of weighted shifts on $B$.
          That is, $T=UD=\sum_{i}T_{i}P_{i}$ where $U=\sum_{i}U_{i}$ is a
          direct sum of shifts (bilateral, unilateral, adjoint of a
          unilateral shift, finite, or cyclic) and $D$ is pure discrete
          with positive real eigenvalues and eigenstates in $B$.  These
          types of shifts correspond respectively to the associated paths
          being two-way infinite, one-way infinite from either end ($m\geq
          0$ or $m\leq 0$) finite with distinct ends, or cyclic.    One
          also has $U=\sum_{l,s}W_{l,s}$.

          For each $i$ $T_{i}=U_{i}D$ is a weighted shift on a subset
          $B_{i}$ of $B$ which spans a subspace $P_{i}\cal H$ of $\cal H$
          \cite{Halmos}.  The projection operators $P_{i}$ are all pairwise
          orthogonal.  More explicitly if $\{\vert m,i\rangle \}$ is the
          basis $B_{i}$ and $U_{i}\vert m,i\rangle \neq 0$, $T_{i}\vert
          m,i\rangle=D_{i,m}\vert m+1,i\rangle$.  $D_{i,m}$ is the
          eigenvalue of $D$ associated to the eigenstate $\vert
          m,i\rangle$. For the adjoint $T^{\dag}\vert m,i\rangle =D_{i,m-1}
          \vert m-1,i\rangle$ if  $U^{\dag}_{i}\vert m,i\rangle \neq 0$. 

          These results, expressed in the computation basis labels, give
          for any state which is not a terminal path state, 
          \begin{eqnarray}
          T\vert l_{m},j_{m},S_{m}\rangle & = & D_{l_{m},S_{m}(j_{m})}\vert
          l_{m+1},j_{m+1}, S_{m+1} \rangle \nonumber \\
          T^{\dag}\vert l_{m},j_{m},S_{m}\rangle & = & D_{l_{m-1},j_{m-
          1},S_{m-1}(j_{m-1})}\vert l_{m-1},j_{m-1},S_{m-1}\rangle.
          \label{TTdagcomp}
          \end{eqnarray}
          These equations use the requirement that $D$ be a local operator
          in that its values depend only on the state of the qubit at the
          head location.  It is independent of the states of qubits at
          other lattice sites.  In this case one has $D\vert
          l,j,S\rangle=D_{l,j,S}\vert l,j,S\rangle=D_{l,S(j)}\vert
          l,j,S\rangle$.  The connection between the eigenvalues of $D$ and
          the values of $\gamma_{l,s}$ is then given directly by
          $D_{l,S(j)}=\gamma_{l,S(j)}$. 


          If $T$ and $T^{\dag}$ are distinct path generating on $B$ then
          the direct sum decomposition of $T$ means that the Feynman
          Hamiltonian \cite{Feynman} of Eq. \ref{ham} is also decomposable
          as $H=\sum_{i}H_{i}P_{i}$ where for each $i$, 
          \begin{equation}
           H_{i}=K(2-T_{i}-T^{\dag}_{i}). \label{tbhi}
          \end{equation}
          Since $T_{i}$ is a weighted shift on the path of states in
          $B_{i}$, $H_{i}$ is a tight binding Hamiltonian on $B_{i}$ and
          thus in the subspace $P_{i}\cal H$. 

          As is noted elsewhere \cite{Benioff3} $H_{i}$ can also be written
          in the form
          \begin{equation}
          H_{i} = K(2-U_{i}-U^{\dag}_{i}) +V_{i} \label{tbhiv}
          \end{equation}
          where $K(2-U_{i}-U^{\dag}_{i})$ is, formally,  the kinetic energy
          associated with the evolution of the computation along path $i$. 
          For the physical model used here it is also the kinetic energy of
          head motion on the qubit lattice. (The kinetic energy depends on
          the magnitude but not the direction of the head momentum on the
          lattice.) 

          The potential $V_{i}=K(U_{i}-T_{i}+U^{\dag}_{i}-T^{\dag}_{i})$ is
          a nearest neighbor off-diagonal potential with matrix elements on
          internal path states given by 
          \begin{equation}
          \langle m^{\prime},i\vert V_{i}\vert m,i\rangle = K[(1-
          D_{i,m})\delta_{m^{\prime},m+1}+(1-D_{i,m-
          1})\delta_{m^{\prime},m-1}]. \label{vjj}
          \end{equation}
          Since $\vert m,i\rangle =\vert l_{m},j_{m}, S_{m}\rangle$ and $D$
          is local, $D_{i,m}=\gamma_{l_{m},S_{m}(j_{m})}$.

          In Schr\"{o}dinger evolution under the action of $H$, Eq.
          \ref{ham}, the choice of which $H_{i}$ is active is determined by
          the initial state $\Psi(0)$.  If $\Psi(0)$ is a superposition of
          of computation basis states in just one path $i$, Then $H_{i}$ is
          the only  active component.  If $\Psi(0)$ is a superposition of
          computation basis states in different paths, then all $H_{i}$ are
          active for just those paths $i$ in which $\Psi(0)$ has a nonzero
          component.  Note that in this case the states in the different
          paths evolve coherently.

          The simplest nontrivial choice for $D$ (other than $D=1$) is that
          the $D$ has two  eigenvalues $1,\gamma $ where $\gamma$ is a
          positive real number between $0$ and $1$. This will be
          implemented here by limiting consideration to generalized quantum
          Turing machines (GQTMs) that have a fixed potential associated
          with the read-1 elementary steps only.  All other qubit read
          steps are potential free.  Read-1 steps are defined to be all
          steps $\vert m,i\rangle \rightarrow \vert m+1,i\rangle$ in which
          an elementary step term $T_{l,1}$ is active (i.e. the qubit at the
          location of the head is in state $\vert 1\rangle$).  This is
          accounted for by setting 
          \begin{equation}
          T=\sum_{l}[(\sum_{s\neq 1}W_{l,s})+\gamma W_{l,1}] \label{Tgqtm}
          \end{equation} 

          \section{The Counting GQTM}
          \label{CGQTM}
          \subsection{The Step Operator}
          The counting GQTM serves as an interesting example of the
          foregoing.  Based on Eq. \ref{Tgqtm} the step operator for this
          GQTM is a sum over seven elements or terms (each term includes
          the j-sum): 
          \begin{eqnarray}
          T & = & \sum_{j=-\infty}^{\infty}(Q_{0}P_{0j}uP_{j}
          +wQ_{0}P_{2j}uP_{j}+Q_{1}P_{0j}uP_{j}   \nonumber \\
           & & \mbox{}+wQ_{1}P_{2j}u^{\dag}P_{j} +\gamma
          Q_{2}v_{xj}P_{1j}u^{\dag}P_{j}+w^{\dag}Q_{2}v_{xj}P_{0j}uP_{j}  
          \nonumber \\
           & & \mbox{}+wQ_{2}P_{2j}uP_{j}) \label{Tex}
          \end{eqnarray}
          The projection operators $Q_{l},\,P_{s,j}\,P_{j}$ are for the
          head in state $\vert l\rangle$, the site $j$ qubit in state
          $\vert s\rangle$, and the head at site $j$.  $w$ is a shift mod 3
          on the three head states ($wQ_{m}=Q_{m+1}w \bmod 3$) and $u$
          shifts the head along the lattice by one site
          ($uP_{j}=P_{j+1}u$).  The need for markers is accounted for here
          by choosing the qubits in the lattice to be ternary with states
          $\vert 0\rangle ,\vert 1\rangle ,\vert 2\rangle$. $\vert
          2\rangle$ is used as a marker and $\vert 0\rangle ,\vert
          1\rangle$ are used for binary strings.  The qubit transformation
          operator $v_{xj}=\sigma_{xj}(P_{0j}+P_{1j})+P_{2j}$ exchanges the
          states $\vert 0\rangle , \vert 1\rangle$ and does nothing to the
          state $\vert 2\rangle$ for the site $j$ qubit. 

          The adjoint $T^{\dag}$ is defined from $T$ in the usual way
          noting that the operators for the head states, head position
          states, and qubit states commute with one another.  Note that
          $v_{xj}^{\dag}=v_{xj}$.  This gives
          \begin{eqnarray}
          T^{\dag} & = & \sum_{j=-\infty}^{\infty}(Q_{0}P_{0j}P_{j}u^{\dag}
          +Q_{0}w^{\dag}P_{2j}P_{j}u^{\dag}+Q_{1}P_{0j}P_{j}u^{\dag}  
          \nonumber \\
           & & \mbox{}+Q_{1}w^{\dag}P_{2j}P_{j}u +\gamma
          Q_{2}P_{1j}v_{xj}P_{j}u+Q_{2}wP_{0j}v_{xj}P_{j}u^{\dag}  
          \nonumber \\
           & & \mbox{}+Q_{2}w^{\dag}P_{2j}P_{j}u^{\dag}) \label{Texdag}
          \end{eqnarray}

          The terms in $T$ have been chosen so that, besides carrying out
          the desired operations, $T$ is distinct path generating in the
          computation basis.  As defined $T$ and $T^{\dag}$ satisfy a
          sufficient condition for this \cite{Benioff}, namely, that 
          $T^{\dag}T=\sum_{t=1}^{7} t^{\dag}t$ and $TT^{\dag}=
          \sum_{t=1}^{7}tt^{\dag}$ where $t$ and $t^{\dag}$ denote terms 
          of $T$ and $T^{\dag}$ in Eqs. \ref{Tex} and \ref{Texdag}.

          For this example all steps are potential free except those in
          which the 5th term is active.  For these a potential, given by
          Eq. \ref{vjj}, is present as $D_{l,1}=\gamma$ for $l=1$.  The
          distribution of potential and potential free steps along a path
          depends on which of the 7 terms is active for a path state.  The
          potential width at any path location is given by the number of
          successive iterations for which the 5th term is active.

          Comparison with Eq. \ref{Tgqtm} shows that for this GQTM ($T=UD$)
          $U$ is a also sum of 7 terms with $U_{2,1}$ the fifth term of Eq.
          \ref{Tex} with $\gamma$ excluded. The other 6 terms of $U$ in the
          $l,s$ sum are given directly by Eq. \ref{Tex}.

          To apply this to a specific example, consider the initial state,
          shown in Figure 1, with the head in state $\vert 0\rangle$ in a
          wave packet localized to the left of the origin.  All qubits are
          in state $\vert 0\rangle$ except those at sites $0,n+1$ which are
          in state $\vert 2\rangle$.  The initial head and lattice qubit
          state are the path labels for this example.  Iteration of $T$ on
          this state generates in turn all the integers as binary strings
          of length $\leq n$.  When the space between the two markers is
          completely filled with $1s$, corresponding to the integer $2^{n}-
          1$, the last pass of the head changes all $1s$ to $0s$. The head
          in state $\vert 1\rangle $ then moves to the right away from the
          marker region as the enumeration is completed.  A more detailed
          description of this process, based on iteration of $T$, is  given
          elsewhere \cite{Benioff3}. 

          \subsection{Potential Distributions}
          The path potential distribution for the counting GQTM is closely
related to the distribution of read-1 steps in all the steps (read-1, read-
          0, and read-2) obtained by iteration of $T$.  This distribution
          can be represented by a function $\cal R$ from the nonnegative
          integers to $\{0,1\}$ where ${\cal R} (j)=1 [0]$ if the $jth$ step
          (or $T$ iteration) is a read-1 [read-0 or read-2] step.  The
          initial state or origin is assumed to be with the head in state
          $\vert 1\rangle$ at position $n+1$. The location of the origin is
          not important provided all read-1 steps are included.  

          The read-1 steps occur during a stepwise search, starting
          from the units place, for the first $0$ after a string of $1s$.
Based on this $\cal R$ would be expected to be related to the function $R$
          from the nonnegative integers to the nonnegative integers such
          that, in the binary string for $j$, $R(j)$ gives the number of $1s$
	  occurring before the first $0$.  Because of this, the properties 
	  of $R$ can be used to obtain a useful expression for $\cal R$.  

The range sequence of $R$, which is $0,1,0,2,0,1,0,3,\cdots$ , is known as
          the hierarchical sequence \cite{Turban}.  It can be expressed
          recursively \cite{Benioff3} by the pair of recursion equations;
          $R_{N}=S_{N-1}N,\; S_{N}=R_{N}S_{N-1}$ for $N=1,2,\cdots$ with
          $S_{0}=0$.  $R_{N}$ is the first $2^{N}$ elements of $R$.  $R$
          is a generalized substitution sequence under the
          substitution rule $n\rightarrow 0n+1$ for $n=0,1,2,\cdots$. The
          generalization denotes the fact that, contrary to the literature
          definitions \cite{BoGh1,KoNo,Queff}, the alphabet is infinite.

          The recursion relations for $R$ also hold for $\cal R$ if $N$ is
          replaced with $\underline{N}$ where $\underline{N}$ denotes the
          string of numbers $0 1^{N}0^{N+1}$.  Here $1^{N}$ and $0^{N+1}$
          denote strings of $N\: 1s$ and $N+1\: 0s$.  That is, for
$N=1,2,\cdots$
          \begin{equation}
          {\cal R}_{N}={\cal S}_{N-1}\underline{N};\: {\cal S}_{N}={\cal
          R}_{N}{\cal S}_{N-1}. \label{calRrecur}
          \end{equation}
where ${\cal S}_{0} =\underline{0}$.

          This equation gives the distribution of read-1 steps in all steps
           associated with enumerating the first $2^{N}$
          integers. An example for $N=6$ is given elsewhere
          \cite{Benioff3}.  The explicit locations of bands of read-1 steps 
     can be obtained by noting that the number of elements in $\cal{S}_{N}$
          and $\cal{R}_{N}$ is given respectively by $2^{N+3}-2(N+3)$ and
          $2^{N+2}-2$.  The number of read-1 bands is equal to $2^{N-1}$.

The relation of the read-1 step distribution to the potential distribution in
the tight binding Hamiltonian, Eq.
          \ref{tbhiv}, for the path corresponding to enumerating the first
          $2^{N}$ integers, can be obtained from  $\cal{R}_{N}$.  Extend
${\cal R}_{N}$ to a two way infinite sequence by adding $0s$ to both ends. 
Define a function $D_{i}$ on the spectrum of $D$ by $D_{i,m}=D_{i}(m)$.
      The function $D_{i}$ is related to the extended ${\cal R}_{N}$ by 
          \begin{equation}
          D_{i}=1+(\gamma -1) {\cal R}_{N}. 
          \end{equation}.
          From this the matrix elements of $V_{i}$ , Eq. \ref{vjj}, are
          given by
          \begin{equation}
          \langle m^{\prime}\vert V_{i}\vert m\rangle = K(1-\gamma)[{\cal
          R}_{N}(m)\delta_{m^{\prime},m+1}+{\cal R}_{N}(m-
          1)\delta_{m^{\prime},m-1}]. \label{vexjj}
          \end{equation}

Eq. \ref{vexjj} shows that the off-diagonal potential barrier associated
with a band of
$n\; 1s$ at path sites $a, \cdots ,a+n-1$ with $0s$ on both sides (i.e.
${\cal R}_{N}(j) = 1$ if $a\leq j\leq a+n-1$ and ${\cal R}_{N}(a-1)={\cal
R}_{N}(a+n) =0$)  extends
from sites $a$ to $a+n$. The potential barrier has a core of $n-1$ sites of
height $2K(1-\gamma)$ flanked by  single site potentials of height $K(1-\gamma)$
at sites $a$ and $a+n$. This is the case whether these sites correspond
to initial or final states in the matrix
elements of Eq. \ref{vexjj}. Note that a potential is associated with the site
$a+n$ for which ${\cal R}_{N}$ has value $0$.  If $n=1$ the barrier is two
sites wide and has a height of $K(1-\gamma)$ as no core is present. The
potential distribution given by Eq. \ref{vexjj} for $N=6$ is shown in Figure 2.

Figure 2 is a good illustration of the general property of the distribution,
that  barriers that are two sites wide ($n=1$) are the most common as $1/2$
the barriers are of this type. One half the remaining barriers are 3 sites
wide $n=2$, and $1/2$ those remaining are 4 sites wide ($n=3$), etc..  

          There are many other possible initial states and paths with
          associated tight binding Hamiltonians besides that given in
          Figure 1. Discussion of these, and the collection of paths into
          equivalence classes, is discussed elsewhere \cite{Benioff3}.

          \subsection{Transfer Matrix}
          Following Erdos and Herndon \cite{Erdos}, solutions for tight
          binding Hamiltonians, such as that given by Eq. \ref{tbhiv}, can
          be written as $\Psi=\Psi_{I}+\Psi_{II}+\Psi_{III}$ where
          $\Psi_{I}$ and $\Psi_{III}$ are the states coresponding to the
          path boundary potential free regions and $\Psi_{II}$ is the state
          in the path region containing the potentials.  For example if the
          path regions for $j< 0$ and $j>n$ are potential free (integers
          label path states) then
          \begin{eqnarray}
          \Psi_{I} & = & \sum_{j=-\infty}^{-1}(Ae^{ikj}+Be^{-ikj})\vert
          j\rangle \nonumber \\
          \Psi_{III} & = & \sum_{j=n+1}^{\infty}(Fe^{ikj}+Ge^{-ikj})\vert
          j\rangle \label{PsiIMF}
          \end{eqnarray}
          Here $j$ is the path state label and $k$ is the momentum in the
          potential free regions.  

          The form of $\Psi_{II}$ depends on the form of the potential.  For
a single barrier corresponding to a band of $N$ read-1 sites extending from
site $a$ to $a+N-1$, 
\begin{eqnarray}
          \Psi_{II}  & = & (C_{1}e^{iha}+D_{1}e^{-iha})\vert a\rangle 
	  +\sum_{j=a+1}^{a+N-1}(C_{2}e^{ilj}+D_{2}e^{-ilj})\vert j\rangle
\nonumber \\
  & & \mbox{}+(C_{3}e^{ih(a+N)}+D_{3}e^{-ih(a+N)})\vert a+N\rangle. \label{psi}
          \end{eqnarray}
If the energy is below the barrier height of $2K(1-\gamma)$, which is the
case of interest here, then $l$ is replaced by $il$.  If the energy is below
$K(1-\gamma)$ then the momentum $h$ is replaced by $ih$. If several potentials are
present $\Psi_{II}$ is a sum of terms of the form of Eq. \ref{psi}.

          The four complex coefficients $A,B,F,G$ are completely determined
          by boundary conditions for the case under study and the
          properties of the transfer matrix.  For example if initial states
          are required to be localized in region I (Figure 1), then $G=0$
          as there is no incoming head motion in region III.  

          The transfer matrix $Z$ relates the coefficients $F,G$ to $A,B$
          according to \cite{Erdos}
          \[ \left| \begin{array}{cc}
          F \\ G \end{array} \right) =Z \left| \begin{array}{cc}
          A \\ B \end{array} \right). \]
          $Z$ is a unimodular (determinant=1) $2\times 2$ matrix which also
          satisfies \cite{Erdos}
          \begin{eqnarray}
          Z_{11} & = & Z^{*}_{22} \nonumber \\
          Z_{12} & = & Z^{*}_{21}. \label{Zprop}
          \end{eqnarray}
 
          For a string of potentials the transfer matrix connecting the
          state on the right to the state on the left is the product of the
          transfer matrices for each of the potentials
          \cite{Erdos,Roy1,Roy2}.  If the potential distribution is given
          by a recursion relation, such as that of Eq. \ref{calRrecur},
          then the transfer matrix $Z_{N}$ for the potential distribution
          corresponding to ${\cal R}_{N}$ is given by the recursion
          relations
          \begin{eqnarray}
          Z_{N} & = & W_{N}X_{N-1} \nonumber \\
          X_{N} & = & X_{N-1}W_{N}X_{N-1} \label{Zrecur}
          \end{eqnarray}
          with $X_{0}|_{11} =e^{2ik}$ and $X_{0}|_{12}=0$. Eq. \ref{Zprop}
          holds for $X_{0}$.  $W_{N}$ is the transfer matrix for the
          potential corresponding to the sequence $\underline{N}
          =01^{N}0^{N+1}$.  As noted earlier this corresponds to a potential
barrier with a core width of $N-1$ sites flanked on each side by a one-site
potential half the core height. Potential free regions of $1$ site and $N+1$
sites surround the barrier.  Note that the order of matrix multiplication in
Eq. \ref{Zrecur} is the inverse of the order in which terms appear in the 
sequence of Eq. \ref{calRrecur}.  

          This use of recursion relations has the advantage that in order
          to obtain the elements of the matrix $Z_{N}$, only polynomially
          many matrix multiplications are needed.  If $Z_{N}$ is obtained
          from the matrices associated with each potential barrier in the
          distribution, then exponentially many matrix multiplications are
          required.

          From Eq. \ref{Zrecur} one sees that the only matrices needed in
          explicit form (other than $X_{0}$) are the $W_{m}$ for $m \leq
          N$. These can be obtained explicitly (see also \cite{Erdos}) by
          setting $(E-H_{i})\Psi =0$ where $H_{i}$ is given by Eqs.
          \ref{tbhiv} and \ref{vexjj}, and the state $\Psi$ is given by
          Eqs. \ref{PsiIMF} and \ref{psi}.  A set of linear equations is 
	  obtained by setting the coefficients of each state $\vert 
	  j,i\rangle$ equal to $0$. that is 
          \begin{equation}
          \sum_{j^{\prime}}\langle j,i\vert E-H_{i}\vert \label{EHj}
          j^{\prime},i\rangle \langle j^{\prime}\vert \Psi \rangle =0.
          \end{equation}
          Because $H_{i}$ contains nearest neighbor interactions only, just
          3 terms in the $j^{\prime}$ sum, $j^{\prime}=j,j\pm 1$
          contribute.

          The linear equations so obtained can be used to derive the matrix
          elements of $W_{m}$ for $\underline{m}=01^{m}0^{m+1}$.  A brief
summary of the derivation is given in the Appendix.  For 
	  energies below the potential core height 
	  (i.e. less than $2K(1-\gamma)$), they are
          \begin{eqnarray}
           W_{m}|_{11} & = & \frac{e^{ik(m+2)}}{2i\gamma \sin k \sinh l}
	   [e^{2ik}\sinh{lm} -2\gamma e^{ik}\sinh{l(m-1)} 
	   +\gamma^{2}\sinh{l(m-2)}] \nonumber \\
          W_{m}|_{12} & = & \frac{e^{ikm}}{2i\gamma \sin k\sinh l}
	  [\sinh{lm} -2\gamma\sinh{l(m-1)}\cos k +\gamma^{2} \sinh{l(m-2)}]
	  \label{Wmat}
          \end{eqnarray}
          where the momenta $l,k$ are related by $2K(1-\cos k)=E
	  =2K(1-\gamma \cosh l)$. For the flank potentials or those for
$m=1$, $E=K(2-(\gamma^{2} +2\gamma \cos{2h} +1)^{1/2})$.  However, the momentum
$h$, Eq. \ref{psi},
does not appear in the above. These equations can be shown by explicit calculation to
hold for $m=1$ and $m=2$ as well as for larger $m$. One also has
$W_{m}|_{11}=W^{*}_{m}|_{22},\: W_{m}|_{21}
          =W^{*}_{m}|_{12}$; and $W_{m}$ is unimodular.

          \section{Energy Bands and Gaps}
          \label{SP}

          As is well known \cite{Others,Roy1,Roy2} the distributions of
          energy bands and gaps for the infinite 1D periodic system, with a
          unit cell satisfying the problem being considered, can be
          obtained from a plot of $Tr Z_{N}$ as a function of the momentum
          $k$. Regions for which $|TrZ_{N}|\leq 2$ correspond to energy
          bands; regions with $|TrZ_{N}|>2$ correspond to energy gaps.  As
          the main interest in the paper is in bound states, $k$ is
          restricted to be less than $\arccos \gamma$. The infinite
          periodic system is referred to as the associated periodic system.

          Test calculations show that except for values of $\gamma$ very
          close to $1$, the values of $Tr Z_{N}$ fluctuate extremely
          rapidly between astronomically large positive and negative values
          (of the order of $\pm 10^{M}$ where $M$ is very dependent on
          $\gamma$ and $N$. For values of $\gamma$ appreciably different
          from $1$ and values of $N$ of order $10$, $M$ has values greater
          than $200$. It is thus necessary to determine values of the two
          free parameters $K,\gamma$ that are reasonable physically. To do
          this it is necessary to assign a spacing $\Delta$ to the lattice
          qubits.  Then the dimensionless momenta $l,k$  are related to the
          usual momenta $l^{\prime},k^{\prime}$ by
          $k=k^{\prime}\Delta,\;l=l^{\prime}\Delta$. 
          The form of $K$ is obtained from the fact that if $T=Y$, the
          bilateral shift on the lattice, then $H$, Eq. \ref{ham}, is the
          symmetrized form of the second derivative on the lattice.  This
          gives 
          \begin{equation}
          K=\frac{\hbar^{2}c^{2}}{2m\Delta^{2}}
          \end{equation}
          where $m$ is the mass of the head in energy units ($mc^{2}$).

          Since  $V=2K(1-\gamma )$, with $\gamma =1-\delta$,
          \begin{equation}
          \delta = \frac{2mV\Delta^{2}}{\hbar^{2}c^{2}}. \label{delta}
          \end{equation}
          For electron systems, $m$ is of the order of the electron mass,
          $V$ is a few electron volts, and $\Delta$ is measured in
          Angstroms.  Taking $m$ equal to 2 electron masses, $\Delta = 1
          \AA$, and $V=2ev$ gives $\delta \simeq 0.001$ or $\gamma \simeq
          0.999$

          The method used for calculation of $TrZ_{N}$ consists of using
          the recursion relations, Eq. \ref{Zrecur}, to obtain $Z_{N}$ and
          then taking the trace.  The well known method of using trace maps
          \cite{KoNo,BoGh2,AvBeGl}, is not used because it is more
          efficient to obtain directly the matrix $Z_{N}$ and then take the
          trace rather than by use of the trace map relations.  In
          particular, obtaining $Z_{N}$ directly from Eq. \ref{Zrecur},
          using Eq. \ref{Zprop}, and then taking the trace requires a
          number of scalar multiplications proportional to $N$.  Use of the
          trace map requires a number of scalar multiplications
          proportional to $N^{2}$.  

          Energy bands and gaps were calculated as a function of the
          dimensionless momentum $k$ for several different values of $
          \gamma$ and $N$.   The results for $\gamma =0.999$ are given in
          Figure 3 for $N=10,\;15,\;18$ as three bands of vertical lines
          connecting horizontal line segments or points.  The upper or
          lower horizontal line segments or points represent respectively
          energy bands or energy gaps. Points represent bands or gaps whose
          width is of the order of the values of the increment or spacing
          $\Delta k$ used to make the calculations.  
	  
          The range of momentum shown is from $k=0.0223$ to $k=\arccos
          \gamma$.  The lower limit is chosen because, to the accuracy of
the calculations, there are no energy
          bands in the region $k < 0.0223$  in which $Tr Z_{N}$ increases
          rapidly to very large values. Each band shows the result of
          calculations using about $2800$ equally spaced grid points which
          gives $\Delta k \approx \gamma /5600$. 

          The lower band for $N=10$ shows that most of the momentum region
          $0.0223 \leq k \leq \arccos \gamma =0.0447$ is covered by energy
          bands separated in most cases by very narrow gaps.  The widest
          gap is at $k\simeq 0.033$ with smaller gaps at $k\simeq
          0.025,\;0.029,\;0.038$, and $0.043$.  

          The middle band for $N=15$ shows that the wide energy bands for
          $N=10$ are split into many smaller bands with intervening gaps. 
          Gaps which were already present at $N=10$ continue to be present. 
          This splitting continues for increasing $N$ as is shown by the
          upper band for $N=18$.  The width and positions of the larger
          gaps at $k\simeq 0.043,\; 0.038,\; 0.033,\; 0.029,\; 0.025$ are
          maintained.  The bands also show that as $N$ increases the number
          of bands increases greatly. However the band widths decrease even
          faster so that the fraction of the momentum region associated
          with energy bands, although sitll large, decreases as $N$
          increases.  

          The distribution of bands or band density is shown by the spacing
          or density of the vertical lines. For example for $N=18$ the
          density is high in the dark regions of the band, such as around
          $k\approx 0.024$ and $0.036$. It is low in the regions of the
          wide gaps.

          Calculations were not made for values of $N>18$ because the range
          and fluctuation of values of $TrZ_{N}$ becomes so extreme that
          the accuracy of calculations that can be made with a reasonable
          amount of time and effort becomes questionable.  Even at $N=18$
          high resolution calculations (to be discussed later) over small
          momentum regions are needed to resolve individual bands.  Recall
          that the results shown  give the energy band and gap
          structure for an infinite periodic structure with a unit cell of
          $2^{N-1}$ potentials with a width and location  distribution
          given by Eq. \ref{calRrecur}.  Thus for $N=18$, about $128,000$
          potentials are present in the unit cell.

          In all figures except Figure 1, the values of the momentum used 
	  in the abcissa are larger than the
actual values by a factor of 1000.  They are also dimensionless.
          This has the advantage that the bands in Figure 3 are valid for
          any values of $m,\; V,\; \Delta$ that give $\delta =0.001$ (Eq.
          \ref{delta}) or $\gamma =0.999$.  One set is the values chosen
          for the electron range.  Another set in the nuclear range
          (probably impossible to realize) that gives $\delta \approx
          0.001$  is $m$ equal to the proton mass, $V=1Mev$ and $\Delta =3
          F$. ($1F=10^{-13}$cm). For the values chosen in the electron
          range the momentum range shown in the figure corresponds to an
          energy range of from $1$ to $2ev$.  

          It is useful to explore a wider variation of parameters, such as
          larger values of $m,\; \Delta$, or $V$  which give different
          values of $\gamma$.  Choosing values which increase values of
          $\delta$ gives one more latitude in a choice of physical model
          parameters.  For this reason calculations were also made for
          $\delta$ larger by a factor of $10$. The results are shown in
          Figure 4 for $\gamma =0.99$ for the same values of $N$ used in
          Figure 3. The momentum range is cut off at $k=0.069$ as there do
not appear to be any bands below the cutoff.

          Compared to the results for $\gamma =0.999$, Figure 4 shows a
          great increase in the fraction of the momentum region occupied by
          energy gaps.  There are large gaps around values of $k=0.084$ and
          $0.10$ and especially $0.12$.  There are many more energy bands
          present than are shown in Figure 3 for the same value of $N$. But
          the widths are so much smaller that the fraction of the region
          alloted to energy bands is much less.  This is especially
          pronounced in the bands for $N=15$ and $N=18$ where energy bands
          show as points only.  For these values of $N$, the maximum width
          of any energy band is only about $0.0002$.  The calculations for
          $N=15$ and $N=18$ should only be taken as giving a rough
          indication of the bands present.  Use of finer grids in the
          calculations would be expected to show more bands of very small
          widths in the high density band regions than are shown in the
          figure.  The large gaps would be expected to remain. 

          Increasing the value of $\delta$ by another factor of $10$ to
          give $\gamma =0.9$ gives the results shown in Figure 5.  They
          were obtained using a grid with over 7000 equally spaced points
with a low momentum cutoff at $0.166$.
          The values of $N$ used are smaller than those in the preceding
          figures because the fluctuations in the values of $Tr Z_{N}$ are
          so large. 

	  The calculations show that very few energy bands are
          present and except for $N=6$ the widths are pointlike.  For $N=6$
          the band grouping appears to be  hierararchical in that pairs of
          doublets  are grouped into pairs of quartets which are in turn
            grouped into  octets.  The appearance of the band at $k\simeq 
	    0.196$ as a singlet instead of a doublet may be a
consequence of the finite resolution of the calculations.  The gaps
          between the groups increase as the group size increases.  There
          are indications that this structure may be preserved for $N=8$
          and possibly for $N=10$.  However, each band splits up into many
          smaller bands as $N$ increases.  More detailed calculations with
          extremely fine grids would be needed to check this.

	  The reason for the low momentum cutoffs in the appearance of bands is 
	  not clear.  They do not appear to be associated with the height of 
	  the narrowest barriers (1/2 of the total),i.e. those with height 
	  $K(1-\gamma)$.   For these barriers the
corresponding values of the momentum  at which the energy equals the barrier
height ($\cos k=(1+\gamma)/2$) are $0.0316,\;0.100,\;0.317$ for
$\gamma=0.999,\;0.99,\;0.9$ respectively.  These values are appreciably
higher than the values at which bands first appear.

          \section{Landauer Resistance} \label{LR}
          Interest in the Landauer Resistance \cite{Landauer,Erdos},
          defined as the ratio of the reflection probability to the
          transmission probability through a finite sequence of potentials,
          is based on the fact that for 1D electronic systems, it is a
          measure of the transmission of the electron through the system.  

          From the standpoint of this paper the Landauer Resistance (LR) is
          of interest because the reciprocal, $1/(1+LR)$, is the
          probability of transmission through the potential sequence.  For
          any path of states for a GQTM, the reciprocal is a measure of the
          probability of completion of the quantum computation as it refers
          to the  transmission of a wave packet from one end of the path to
          the other through potentials associated with the path states. 
          For the counting GQTM it is a measure of the probability of
          completion of the counting with potentials associated with read-1
          steps only. 

          Landauer \cite{Land1} has emphasized that reflection from
          background potentials associated with each step of a quantum
          computation degrade the computation due to multiple reflections
          from and transmissions through the potentials. In particular for
          energies below the potential height the transmission would be
          expected to decay exponentially as the packet must tunnel through
          each of the potentials, 

          There are indications that this may not be true for potential
          sequences which are not random.  For example it has been shown
          \cite{Roy1,Roy2} that for Thue-Morse, Fibonnaci, and periodic
          sequences in the Kronig-Penney model, the Landauer Resistance 
          fluctuates rapidly with energy between very high values $>>1$ and
          very low values $<<1$.  In addition high values are associated
          with energy gaps in the infinite associated periodic system; low
          values are associated with energy bands.

          For these reasons the Landauer Resistance (LR) has been
          calculated for the counting GQTM for various values of $N$ and
          $\gamma$ as a function of the dimensionless momentum $k$.   The
          calculations are  done using the result that the LR is given by
          \cite{Erdos}
          \begin{equation}
          LR= |Z_{N}|_{12}|^{2} 
          \end{equation}
          where $Z_{N}$ is given by Eq. \ref{Zrecur}.  The results are
          shown in figures that show the functional dependence of the log
          (base 10) of the Landauer Resistance (log LR) on the momentum.
          The momentum units are dimensionless. 

          The results for $\gamma =0.999$ and $N=10$ are shown in Figure 6.
          The enery band and gap structure for the associated periodic
          system is reproduced at the bottom of the figure.  The
          calculations show a rapidly fluctuating $\log LR$ with large
          maxima at momentum values corresponding to energy gaps.   The
          heights of these large maxima appear to be correlated with
          associated gap widths; wide gaps have greater maxima than
          narrower ones.   Also the general trend of the values such as the
          height of the maxima, is towards lower values as the momentum
increases.  Both
          these features were noted by Roy and Khan \cite{Roy1,Roy2}. 
          However for the smaller maxima, which occur in regions of energy
          bands of fairly uniform width separated by narrow gaps, the
          opposite appears to be the case in that the maxima appear to be
          associated with energy bands.  In addition $\log LR$ has sharp
          spikes to very low values at the regions of these narrow gaps.  

          An interesting feature of the results is that for most values of
          the momentum $\log LR$ is less than $0$.  A  value of $0$
          corresponds to a $50\%$ probability of transmission through the
          potentials.  Thus for this case the probability of completion of
          the quantum computation is $>50\%$ for most momenta.  For
          selected values the probability is close to $100\%$.  These low
          values are very likely correlated with the result that most
          momenta are in energy bands with only a small fraction of the
          momentum region taken up with gaps.

          Figure 7 shows a higher resolution plot of $\log LR$ for a
          portion of the momentum region.  It shows very clearly the
          association of the sharp minima to band edges.  A study of the
          actual computation points for the figure shows that the minima are
          associated with the upper edge of a band just below a gap and not
          with the lower edge of a band just above a gap.  Also the depth
          of the minima appear to depend on the width of the associated gap
          in that in general the depths are much less for wider gaps than
          for narrower ones.  For example the minimum with the relatively
          wide gap at $k=0.0355$ is higher than the sharp spike minima on
          either side with relatively narrow gaps.  The minimum associated
          with the even wider gap at $k=0.038$ has almost disappeared.

          The results of a calculation of $\log LR$ for the same value of 
          $\gamma$ but for a larger value of $N=18$, are shown in Figure 8. 
          Just as the band spectrum is much more finely divided $\log LR$
          fluctuates much more rapidly and violently.  For the purposes of
          clarity the plot is cut off at values of $\log LR >20$
          ($LR>10^{20}$).  The main reason is that the interest here is in
          the behaviour of $\log LR$ around $0$ and not how high the values
          can go.  

          The results show that there are a great number of values of the
          momentum for which $\log LR$ is $0$ or less.  However there are
          also a large number of fluctuations to much higher values.
          Because of the finely divided nature of the spectrum and extreme
          violence of the fluctuations, the figure should be taken only as
          an approximate indication of the dependence of $\log LR$ on the
          momentum.   

          A higher resolution calculation over the range $0.032\leq k\leq
          0.033$, Figure 9, shows very clearly the preponderance of values
          at or below $0$. as well as the rapid fluctuation to large
          maxima.  The heights of the maxima are clearly correlated with
          the widths of the associated gaps.  However, the individual small
          fluctuations are barely resolved.  The heights of the peaks also
          show a regularity that can be correlated with the widths of the
          potentials in the distribution.  The regularity appears to be
          correlated with the hierarchical structure of the sequence given
          by Eq. \ref{calRrecur}.

          A very high resolution calcuation over a very small momentum
          region of $0.03225\leq k\leq 0.03228$ for $\gamma=0.999\; N=18$
          is shown in Figure 10.  The figure shows very clearly the
          correlation between the decrease in the depths of the minima  and
          the increase in the widths of the associated gaps in the band
          spectra.  For example the gap at $k=0.032261$ is sufficiently
          wide that the associated minimum in $\log LR$ has just about
          disappeared with only a slight dip present.  Shallow
          minima are associated with the narrower gaps at $k=0.032253$ and
          $k=0.032269$.  The figure shows also that band widths are much
          more uniform than gap widths which vary widely.  This point was
          noted before for smaller values of $N$.

           These figures show that there are fairly wide regions of the
momentum where, except for a few sharp spikes, $\log LR \leq 0$.  It is
noteworthy that this corresponds to transmission through more than $128,000$
potential barriers with the energy below the barrier height for 1/2 these
potentials for $0.0316 \leq k\leq \arccos{\gamma}$. For $k\leq 0.0316$ the
energy is below the barrier height for all the potentials.  It is possible
that the high transmission is due to the quasiperiodicity of the potential
distribution; however more work is needed to test this possibility.

          In order to show the dependence of $\log LR$ on $\gamma$ for
          fixed $N$, calculations were made for $N=10$ and $\gamma =0.99$.
          this value is larger by a factor of $10$ than that used so far. 
          It corresponds to the electron regime with larger parameters in
          Eq. \ref{delta} to give $\delta=0.01$.  The results are shown in
          Figure 11.

          The figure shows very clearly the correspondence between maxima
          in $\log LR$ and gaps in the spectrum. Large gaps are associated
          with high peaks and small gaps with small peaks. Also the
          correlation between the peak heights and the hierarchical
          structure of $\cal R$, Eq. \ref{calRrecur}, appears to be
          present. In regions where the band and gap widths are small,
          $\log LR$ fluctuates between values of about $1$ and $-2$ or
          less.  On average the values of $\log LR$, even in small gap
          regions, are higher than for $\gamma =0.999$.   This is to be
          expected as smaller values of $\gamma$ correspond to larger
          values of the potential.   On average the transmissivity is
          appreciable as there are extended regions where the values of
          $\log LR$ are around $0$.

          Calculations for $\delta$ larger by another factor of $10$ to
          $\delta=0.1$ ($\gamma =0.9$) are shown in Figure 12 for $N=8$.  
          The extreme violence of fluctuations in the values of $\log LR$,
          even for the relatively low value of $N=8$ is evident. This is to
          be expected since higher values of $\delta$ corresponds to higher
          values of the potential (with $m$ and $\Delta$ fixed).  Also
          since most of the space is occupied by gaps, $\log LR$ has very
          high values and approaches 0 only in the regions of narrow bands.

          A high resolution calculation for $\gamma =0.9$ and $N=8$ over a 
	  small portion of the
          momentum region $0.40\leq k\leq 0.415$ is shown in Figure 13.  
          The figure shows very clearly the association of minima in the LR
          with upper edges of energy bands and the loss of minima with
          increasing gap width.  The gap including the value of $k=0.408$
          is so wide that the associated minimum in $\log LR$ is completely
          gone.  Also the values of the LR in the small momentum region
shown in the figure are between
          10 and 100 on average with values at or below $1$ ($\log LR \leq
          0$) occuring only rarely. 

          \section{Discussion}

          To summarize, the main characteristics of the band spectra
          shown here include the relative uniformity of band widths
          compared to gap widths over the bound state momentum regions. 
          This is especially apparent in Figure 3.  This figure and Figure 4 
	  suggest that fluctuations in bands widths do appear to increase 
	  somewhat as $\delta$ increases. 
	  
          The figures also show that as $\delta$ increases the gap widths
          and the fraction of momentum range occupied with gaps increases
          greatly.  For $\delta =0.001$ most of the $k$ region is occupied
          by bands.  For $\delta=0.01$ a large fraction is taken up by gaps
          and for $\delta =0.1$ almost all the region is gaps with a few
          very narrow bands present.

          Figures 3-5 show that for fixed values of $\gamma$ as $N$ is
          increased, energy bands are split into increasing numbers of
          bands with decreasing widths. Gaps that exist for a given value
          of $N$ appear to be preserved as $N$ is increased above the given 
          value.  This suggests but does not prove that in the limit of
          $N=\infty$, for each $\gamma$ the spectrum of the tight binding
          Hamiltonain is singular continuous and is a Cantor set.  However,
          the proofs in the literature \cite{BeBoGh,BoGh1,Suto} do
          not hold here as the generalized substitution sequence of Eq.
          \ref{calRrecur} is not primitive.  

          The values of $\log LR$ also fluctuate widely over the momentum
          range with the fluctuations very dependent on the values of
          $\delta$ and $N$.  Sharp minima are located at or near band-gap
          boundaries with more minima located at or near gap-above-band
edges than at or near band-above-gap edges.  The reason for the location of 
minima at or near band gap
          boundaries is not clear. It is worth noting that, based on the
          properties of $Z_{N}$, Eq. \ref{Zprop}, $LR=|Z_{N}|_{1,1}|^{2}-
          1$.  At the boundaries $|Tr Z^{N}/2| =1$.  This gives 
          \begin{equation}
          LR=|\Im (Z_{N}|_{1,1})|^{2}
          \end{equation}
          at the boundaries.  $\Im (x)$ denotes the imaginary part of $x$.

Another aspect of the minima is that none are associated with
          boundaries of sufficiently wide gaps.  The figures show that  in
          almost all cases the depth of the minimum in $\log LR$ is very
          dependent on the width of the associated gap.  If the gap is
          narrow, the minimum is deep.  As the gap becomes wider the
          associated minimum becomes shallower until it disappears for
          sufficiently wide gaps.  An example of this where the minima has
          almost disappeared is the gap just below $k=0.03227$ in Fig 10.
          Other examples can be found in the other figures. 

          One consequence is that for wide gaps the maxima in the LR appear
          to be centered over the gaps. For narrow gaps there is an
          associated  minimum in the LR .  For these cases the maxima in
          the LR appear to be centered over the bands and not the gaps.  

          These results differ from those obtained by Roy and Kahn
          \cite{Roy1,Roy2} for the Thue-Morse lattice in that they found
          all maxima in the LR centered on the gaps irrespective of the gap
          widths.  The reason for this difference is not clear.  The
          results obtained here do agree with those of Roy and Khan in that
          the larger LR maxima are associated with wide gaps, and the
          smaller maxima are associated with narrower gaps.  As noted, here
          the smaller maxima tend to be centered over the bands and not the
          gaps.   

          As is well known the Landauer Resistance is a measure of the
          transmission through a sequence of potentials.  For the case at
          hand it is a measure of the completion probability for the
          counting generalized quantum Turing machine to enumerate the
nonnegative integers up to
           $2^{N}-1$.  However the strong dependence of the
          LR on $k$ means that care must be taken in its use.  For example
          Figure 1 shows an initial state represented as a wave packet over
          different head lattice positions with a fixed lattice qubit
          state.  Transformed to momentum space the initial state is in
          essence a wave packet over the head momentum $k$ with the same
          fixed head internal state and lattice qubit state. That is
          $\Psi=\int _{-\pi}^{\pi}c_{k}\vert l,k,S\rangle dk$ where $\vert
          l,k,S\rangle$ is a state for the head in state $l$ and momentum
          $k$ and the qubit lattice in state $\vert S\rangle$.

          Linearity of the Schrodinger Equation gives the result that, for
          each component state $c_{k}\vert l,k,S\rangle$, the transmission
          and reflection coefficients are given respectively by
          \begin{eqnarray} 
          F & = & \frac{c_{k}}{Z_{N}|_{2,2}} \nonumber \\
          B & = & -\frac{Z_{N}|_{21}c_{k}}{Z_{N}|_{22}} \label{Cnorm}
          \end{eqnarray}
          Here Eqs. \ref{PsiIMF} and \ref{psi}  were used along with the 
	  properties of
          $Z_{N}$, Eq. \ref{Zprop}.  The coefficient $G=0$ as there is no
          incoming component in region III. As the results in Section \ref{LR}
	  show, the coefficients $F$ and $B$ are very dependent on $k$.  Of
          course the LR, given by $LR=|B|^{2}/|F^{2}|$, is independent of
          the normalization to $c_{k}$ shown in Eq. \ref{Cnorm}.

          More general linear superpositions in the initial state are
          possible.  For example the most general initial state for the
          counting GQTM is given by $\Psi =\sum_{l,S}\int_{-\pi}^{\pi}
          c(l,k,S)\vert l,k,S\rangle$.  Each component in the $S$ and $l$
          sums corresponds to a different input state and a different
          quantum computation.  This ability of quantum computers to carry
          out simultaneously computations on different input states i.e.
          parallel quantum computation was first described by Deutsch
          \cite{Deutsch}.

          From the perspective of this paper an interesting aspect of the
          above is that Schr\"{o}dinger evolution with one Hamiltonian, Eq.
          \ref{ham} for the system is equivalent to a linear superposition
          of different Schr\"{o}dinger evolutions each with a different
          tight binding Hamiltonian.  This follows from the fact that in
          general for each component in the $l,S$ sum, the potential
          distribution on the corresponding state path depends on $\vert
          S\rangle$ and on $\vert l\rangle$.  This is an example of a
          linear superposition of the action of component Hamiltonians in
          the decomposition of the Feynman Hamiltonian, Eq, \ref{ham} into
          a sum over tight binding Hamiltonians, Eq. \ref{tbhi}. 

          \section*{Acknowledgements}
          This work is supported by the U.S. Department of Energy, Nuclear 
          Physics Division, under contract W-31-109-ENG-38.

\begin{appendix}
\begin{center}
{\bf APPENDIX}
\end{center}
As noted in the text around Eq. \ref{EHj}, solution of $(E-H)\Psi =0$ gives
a set of linear equations, one for each path state $\vert j\rangle$.  In
regions of constant potential or that are potential free, $\gamma =1$, the
linear equations have the same form:
\begin{equation}
(jXY)(E-2K)+K\gamma (j-1XY)+K\gamma (j+1XY) =0. \label{X}
\end{equation}
Here $(jXY)$ is a shorthand notation for $Xe^{iqj}+Ye^{-iqj}$ where
$X,Y=A,B;\; C_{i},D_{i}$; or $F,G$ respectively of Eqs. \ref{PsiIMF} and
\ref{psi} and q is the appropriate momentum.

Solution of this equation gives $E=2K(1-\gamma \cos q)$.  For a band of $m$
read-1 sites (from $a$ to $a+m-1$) surrounded by $0s$ on both sides, use of
Eqs. \ref{PsiIMF}, \ref{psi}, and \ref{EHj} gives 6 equations different from
Eq. \ref{X} for the sites $a-1,\; a,\; a+1,\; a+m-1,\; a+m,\;a+m+1$. In order of
these sites, they are
\begin{eqnarray*}
(aAB) -(aC_{1}D_{1}) & = & 0 \\
(aC_{1}D_{1})(E-2K) +K((a-1AB)+K\gamma (a+1C_{2}D_{2}) & = & 0 \\
(a+1C_{2}D_{2})(E-2K) +K\gamma (aC_{1}D_{1})+K\gamma (a+2C_{2}D_{2}) & = & 0
\\
(a+m-1C_{2}D_{2})(E-2K)+K\gamma (a+m-2C_{2}D_{2})+K\gamma (a+mC_{3}D_{3}) &
= & 0 \\
(a+mC_{3}D_{3})(E-2K)+K\gamma (a+m-1C_{2}D_{2})+K(a+m+1FG) & = & 0 \\
(a+mFG) -(a+mC_{3}D_{3}) & = & 0. 
\end{eqnarray*}
Here Eq. \ref{X} has been used to simplify the first and last equations to
remove $(a-1AB)(E-2K)$ and $(a+m+1FG)(E-2K)$.

The first equation is used to remove $(aC_{1}D_{1})$ from the next two
equations.  Combination of the resulting two equations with Eq. \ref{X}
gives a $2\times 2$ matrix connecting $(a+1C_{2}D_{2})$ to $(aAB)$.  In a
similar way the second 3 equations are used to give a $2\times 2$ matrix
connecting $(a+mFG)$ to $(a+m-1C_{2}D_{2})$.  The product (taken in the
correct order) of these two matrices and a matrix connecting
$(a+m-1C_{2}D_{2})$ to $(a+1C_{2}D_{2})$ gives the final result.  The matrix
for $W_{m}$ is obtained by setting $a=1$ and translating the result by
another $m+1$ potential free sites.
\end{appendix}

          \newpage

          \begin{center}
          FIGURE CAPTIONS
          \end{center}

          Figure 1.  Initial and Final States for Counting GQTM for the
          First $2^{n}$ Binary Numbers.  All lattice qubits are in state
          $\vert 0\rangle$ except those at sites $0$ and $n+1$ which are in
          state $\vert 2\rangle$.  The initial and final head states are
          shown as wave packets with internal head states $\vert 0\rangle$
          and $\vert 1\rangle$ to the left and right respectively. \\
          \\

	  Figure 2.  The  Potential Distribution for $N=6$.  The potential
height, in units of $K(1-\gamma )$, is plotted against path position.  A
potential height of $1$ or $2$ corresponds respectively to one or both terms
of Eq. \ref{vexjj} being active. \\
\\
          Figure 3.  Energy Band Spectra for the Associated  Periodic
          Systems for $\gamma =0.999$ for three values of N as a Function
          of Momentum $k$. The momentum range is from the least value for
          which $|Tr Z_{N}|\simeq 2$ up to $k=\arccos \gamma$.  Within each
          long band or ribbon associated with each $N$, the upper
          horizontal line segments correspond to energy bands; the lower
          horizontal segments correspond to energy gaps.  Very short bands
          or gaps appear as points.  For $N=15$ and especially $N=18$, band
          and gap density is too high in several momentum regions to be
          resolved at the scale shown. \\
          \\
          Figure 4. Energy Band Spectra for the Associated Periodic System
          for the Same Values of $N$ as were used in Figure 3 but with
          $\gamma =0.99$.  See the caption of Figure 3 for additional
          details.  Note that, compared to the spectra in Figure 3, the
          gaps are much wider and the energy bands are narrower.   \\
          \\
          Figure 5. Energy Band Spectra for the Associated Periodic System
          for $\gamma=0.9$ and $N=6,8,10$. The caption of Figure 3 gives
          more details.  Lower values of $N$ are shown because at the
          higher values of $N$ the fluctuations of $TrZ_{N}$ are so extreme
          that calculations become impractical.  Even for the values of
          $N=8$ and $N=10$ there are no assurances that the grid used (with
          more than 7,000 equally spaced points) is fine enough to capture
          all bands. \\
          \\
          Figure 6.  The Log (base 10) of the Landauer Resistance Plotted
          as a Function of the Momentum for $N=10$ and $\gamma =0.999$. 
          The momentum range is the same as that used in Figure 3. The
          corresponding energy band spectrum for the associated periodic
          system, from Figure 4, is shown at the bottom of the figure. \\
          \\
          Figure 7. An Expanded Plot of the Log (base 10) of the Landauer
          Resistance as a Function of $k$ for $N=10$ and $\gamma =0.999$
          over the Range $0.034\leq k\leq 0.040$. The corresponding
          expanded band spectrum is also shown.  The correspondence between
          minima in $\log LR$ and band-gap boundaries is clearly shown as
          is the decrease in the magnitude of the minima with increasing
          gap width. \\
          \\
          Figure 8. The Log (base 10) of the Landauer Resistance as a
          Function of the Momentum for $N=18$ and $\gamma=0.999$.  The
          corresponding energy band spectrum is also shown.  In order to
          show values of $\log LR$ in the neighborhood of $0$, the ordinate
          values were cut off at 20.  Fluctuations in $\log LR$ are so
          rapid and extreme that individual maxima and minima in the
          neighborhood of $0$ are not resolved. \\
          \\
          Figure 9. Expanded Plot of the Log (base 10) of the Landauer
          Resistance for $N=18$, $\gamma=0.999$ over the range $0.032\leq
          k\leq 0.033$  The corresponding expanded band spectrum is also
          shown.  The range is a small region in Figure 8 showing
          especially rapid fluctuations.  The correlation between the
          height of maxima in $\log LR$ and widths of associated gaps is
          shown clearly. \\
          \\
          Figure 10.  A Very High Resolution Plot of the Log (base 10) of
          the Landauer Resistance for $N=18$, $\gamma=0.999$ Over a Very
          Small Momentum Region of $0.03225\leq k\leq 0.03228$.  The
          expanded band spectrum is also shown.  Individual maxima and
          minima are clearly resolved.  As is the case in Figure 7 for
          $N=10$, the location of minima at or near band-gap boundaries,
          and the decrease in the minima with increasing gap widths are
          clearly shown. \\
          \\
          Figure 11.  The Log (base 10) of the Landauer Resistance Plotted
          as a Function of the Momentum for $N=10$ and $\gamma=0.99$.  The
          corresponding band spectrum is also shown.  The momentum range is
          from the least value of $k$ for which $|TrZ_{N}| \simeq 2$ up to
          $\arccos \gamma$.  High maxima in $\log LR$ are clearly assocated
          with wide gaps. \\
          \\
          Figure 12.  Plot of the Log (base 10) of the Landauer Resistance
          and Corresponding Band Spectrum for $N=8$ and $\gamma=0.9$.  The
          values of $k$ range from the smallest value for which $|Tr Z_{N}|
          \simeq 2$ to $\arccos \gamma$.  The values of $\log LR$ are cut
          off at 20 in order to show the fine structure at smaller values.
          \\
          \\
          Figure 13.  Expanded Plot of the Log (base 10) of the Landauer
          Resistance and Corresponding Band Spectrum for $N=8$ and
          $\gamma=0.9$ for $0.400\leq k\leq 0.415$.  The association
          between minima in $\log LR$ and band-gap boundaries is weaker
          than for values of $\gamma$ closer to $1$ in that some minima are
          close to the middle of gaps.  


\newpage

          \begin{thebibliography}{99}


          \bibitem{Sh}
          D. Shechtman, I. Blech, D. Gratias, and J. Cahn, Phys. Rev.
          Letters {\bf 53} 1951 (1984).

          \bibitem{FuO}
          T. Fujiwara and T. Ogawa (Eds) {\it Quasicrystals}, Proceedings
          of the 112th Taniguchi Symposium, Shima, Mie Prefecture, Japan
          Nov 14-19 1989, Springer Series in Solid State Sciences 93,
          Springer Verlag New York.

          \bibitem{DiVSt}
          D. DiVincenzo and P. Steinhardt Eds., {\it Quasicrystals the
          State of the Art}, in Directions in Condensed Matter Physics-Vol.
          11, World Scientific Singapore 1991.

          \bibitem{BoGh1}
          A. Bovier and J-M. Ghez, J. Phys. A: Math. Gen. {\bf 28} 2313
          (1995).

          \bibitem{defsub}
          Let $A$ be a finite alphabet, $A^{*}$ the set of all finite words
          of $A$, and $A^{N}$ the set of all one way infinite words of $A$.
          A substitution rule replaces each letter $\alpha$ of $A$ with a
          word from $A^{*}$. For each $\alpha$ a sequence of words in
          $A^{*}$ can be generated by starting with $\alpha$ and iterating
          the substitution rule.  The limit sequence in $A^{N}$ is a
          substitution sequence if it is invariant under the substitution.

          \bibitem{Luck}
          J.M. Luck, Phys. Rev. B {\bf 39} 5834, (1989)

          \bibitem{BeBoGh}
          J. Bellissard, A. Bovier, and J-M. Ghez, Commun. Math. Phys. {\bf
          135} 379 (1991)

          \bibitem{Suto}
          A. S\"{u}t\H{o}, Jour Stat. Phys. {\bf 56} 525 (1989).

          \bibitem{BoGh2}
          A. Bovier and J-M. Ghez, Commun. Math. Phys. {\bf 158} 45 (1993);
          {\bf 166} 431 (1994).

          \bibitem{Horn}
          M. H\"{o}rnquist and M. Johansson, J. Phys. A Math. Gen. {\bf 28}
          479 (1995).

          \bibitem{Ryu}
          C. S. Ryu, In-mook Kim, G. Y. Oh, and M. H. Lee, Phys. Rev. B 
          {\bf 49} 14991 (1994).

          \bibitem{AvBeGl}
          Y. Avishai, D. Berend, and D. Glaubman, Phys. Rev. Lett. {\bf 72}
          1842 (1994).

          \bibitem{AvBe1}
          Y. Avishai and D. Berend, Phys. Rev B-II {\bf 43} 6873 (1994).

          \bibitem{AvBe2}
          Y. Avishai and D. Berend, Phys. Rev B-II {\bf 45} 2717 (1992).

          \bibitem{GaKh}
          V. M. Gasparian and A. Gh. Khachatrian, Solid State Commun. {\bf
          85} 1061 (1993).

          \bibitem{Erdos1}
          P. Erdos and R. C. Herndon, Solid State Commun. {\bf 98} 495
          (1996).

          \bibitem{Landauer}
          R. Landauer, Phil. Mag. {\bf 21} 863, (1970).

          \bibitem{Erdos}
          P. Erdos and R. C. Herndon, Adv. Phys, {\bf 31} 65, (1982)

          \bibitem{Roy1}
          C. Roy and A. Khan, Phys. Rev. B, {\bf 49} 14979 (1994): 

          \bibitem{Roy2}
          C. Roy and A. Khan, Solid State Commun. {\bf 92} 241 (1994).

          \bibitem{RoyBasu}
          C. Roy and C. Basu, Phys. Stat. Sol. (b) {\bf 177} 315 (1993).

          \bibitem{Benioff3}
          P. Benioff, To Appear in Phys Rev. Letters

          \bibitem{Land1}
          R. Landauer, Philos. Trans. Roy. Soc. London {\bf A353} 367
          91995); Physics Letters A {\bf 217} 188 (1996); Physics Today
          {\bf 44} No. 5, 23 (1991).


          \bibitem{Peres}
	  A. Peres, Phys. Rev. {\bf A32} 3266 (1985)
	  
	  \bibitem{Zurek}
	  W. H. Zurek, Phys. Rev. Lett. {\bf 53} 391 (1984)
	  
          \bibitem{Benioff1}
          P. Benioff, Jour. Stat. Phys. {\bf 22} 563 (1980); Ann. NY Acad.
          Sci. {\bf 480} 475 (1986).

          \bibitem{Benioff}
          P. Benioff, Phys. Rev A {\bf 54} 1106 (1996).

          \bibitem{Deutsch}
          D. Deutsch, Proc. Roy. Soc. London Ser. A {\bf 400} 97 (1985);
          {\bf 425} 73 (1989).

          \bibitem{Lloyd1}
          S. Lloyd, Sci. American {\bf 273} No.4, 140, (1995); Journal of
          Modern Optics, {\bf 41} 2503 (1994).

          \bibitem{Div}
          D. P. DiVincenzo, Science, {\bf 270} 255 (1995)

          \bibitem{ChLa}
          I.L. Chuang, R. Laflamme, P.W. Shor, and W. H. Zurek, Science
          {\bf 270} 1633 (1995)

          \bibitem{Unr}
          W. G. Unruh, Phys, Rev. A {\bf 51} 992 (1995).

          \bibitem{LaMi}
          R. Laflamme, C. Miguel, J. Pablo Paz, and W. H. Zurek, Phys Rev.
          Letters {bf 77} 198 (1996).

          \bibitem{Feynman1}
          R. Feynman, Int Jour. Theoret. Phys. {\bf 21} 467 (1982).

          \bibitem{Lloyd2}
          S. Lloyd, Science {\bf 273} 1073 (1996); C. Zalka {\em Efficient
          Simulation of Quantum Sysytems by Quantum Computers}, Report No.
          BUTP-96/11, March 25, 1996, Los Alamos Archives quant-ph/9603026.

          \bibitem{Shor}
          P. Shor, in {\it Proceedings of the 35th Annual Symposium on the
          Foundations of Computer Science} (IEEE Computer Society, Los
          Alamitos, CA 1994), p. 124.  For a recent review see A. Ekert and
          R. Jozsa, Rev. Mod. Phys. {\bf 68} 733 (1996).

          \bibitem{Shor1}
          P. W. Shor, Phys. Rev. A {\bf 52} R2493 (1990); R. LaFlamme, C.
          Miquel, J.P. Paz, and W. H. Zurek, Phys. Rev. Letters {\bf 77}
          198 (1996).

          \bibitem{Lloyd}
          S. Lloyd, Phys. Rev. Letters, {bf 75} 346 (1995); D. P.
          DiVencenzo, Phys. Rev. A {\bf 51} 1015 (1995); D. Deutsch, A.
          Barenco, and A. Ekert, Proc. Roy. Soc. London A {\bf 449} 669
          (1995).

          \bibitem{Turban}
          This sequence first appeared in the literature in conjunction
          with work on relaxation of hierarchical systems (B. A. Huberman
          and M. Kerzberg, J. Phys. A: Math. Gen. {\bf 18} L331 (1985)) and
          recently in work on aperiodic Ising chains (F. Igl\'{o}i and L.
          Turban, Los Alamos Archives cond-mat/9606118).  The sequence is
          referred to as the hierarchical sequence.  The author thanks Dr.
          Turban for these references.

          \bibitem{Feynman}
R. Feynman, Optics News, {\bf 11} 11 (1985); reprinted in Foundations of
Physics, {\bf 16} 507 (1986).

          \bibitem{Halmos}
	  P. R. Halmos, {\em A Hilbert Space Problem Book, 2nd Edition},
Springer Verlag, New York, 1982.

          \bibitem{KoNo}
          M. Kol\'{a}\v{r} amd F. Nori, Phys. Rev. B {\bf 42} 1062 (1990)

          \bibitem{Queff}
          M. Queff\'{e}lec, {\it Substitution Dynamical Systems-Spectral
          Analysis}, Lecture Notes in Mathematics 1294, Editors A. Dold and
          B. Eckmann, Springer Verlag, Berlin 1987.

          \bibitem{DiVSt1}
          D. P. DiVicenzo and P. J. Steinhardt, Progress and Current Issues
          in Quasicrystals, in D. P. DiVincenzo and P. J. Steinhardt Eds.,
          {\it Quasicrystals the State of the Art} Directions in Condensed
          Matter Physics-Vol. 11, World Scientific Singapore 1991, pp 1-15.

          \bibitem{Others}
          For example:  M. Kohmoto, L. P. Kadanoff, and C. Tang, Phys. Rev.
          Letters {\bf 50} 1870 (1983); M. Kohmoto, International Jour.
          Modern Phys. {\bf 1} 31 (1987); G. Gumbs and M. K.  Ali, Phys.
          Rev. Letters {\bf 60} 1081 (1988); M. Kolar and M. K. Ali, Phys.
          Rev B {\bf 39} 426 (1989).

          \end{thebibliography}
          
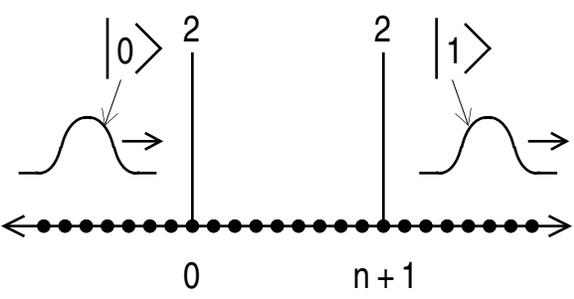
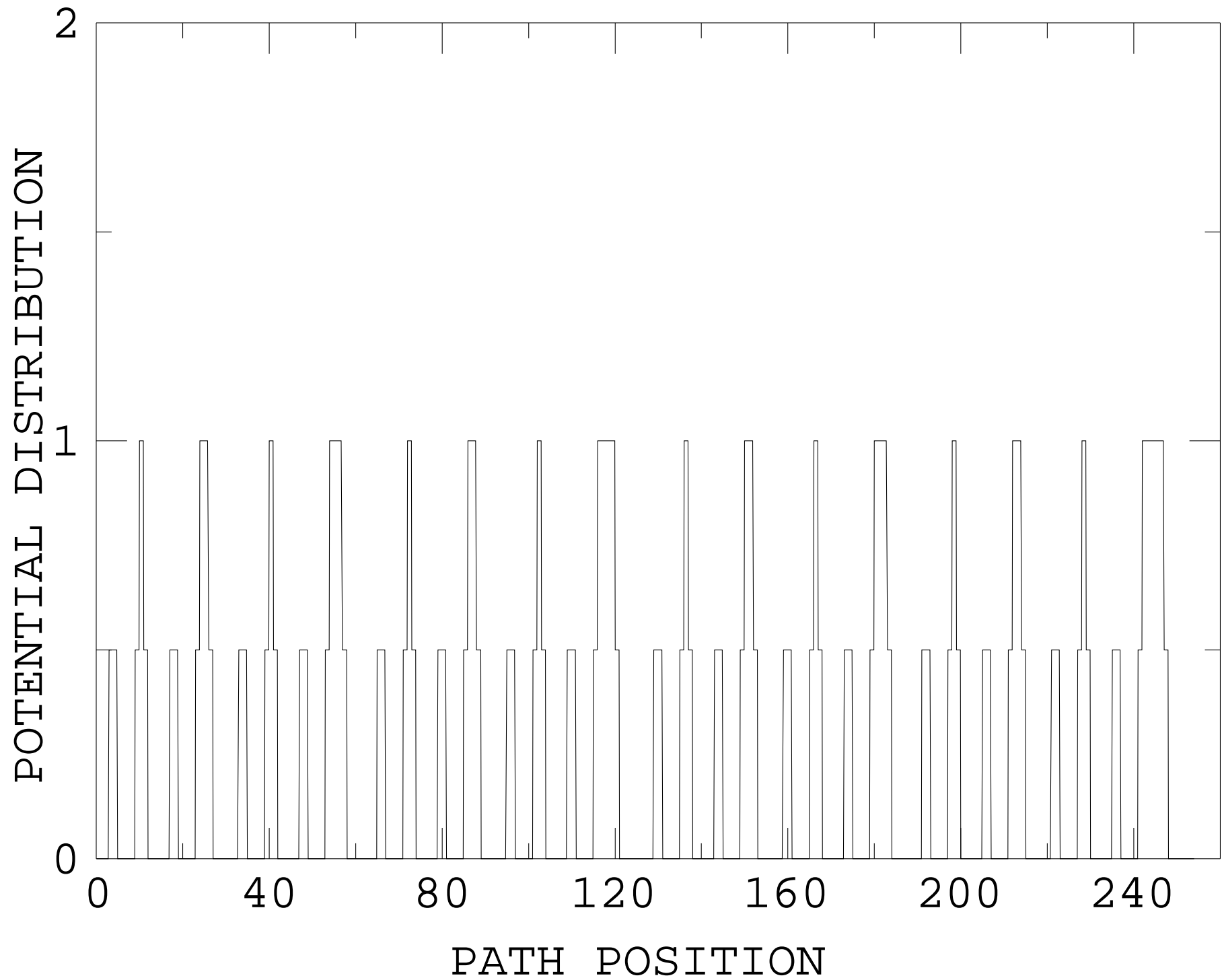
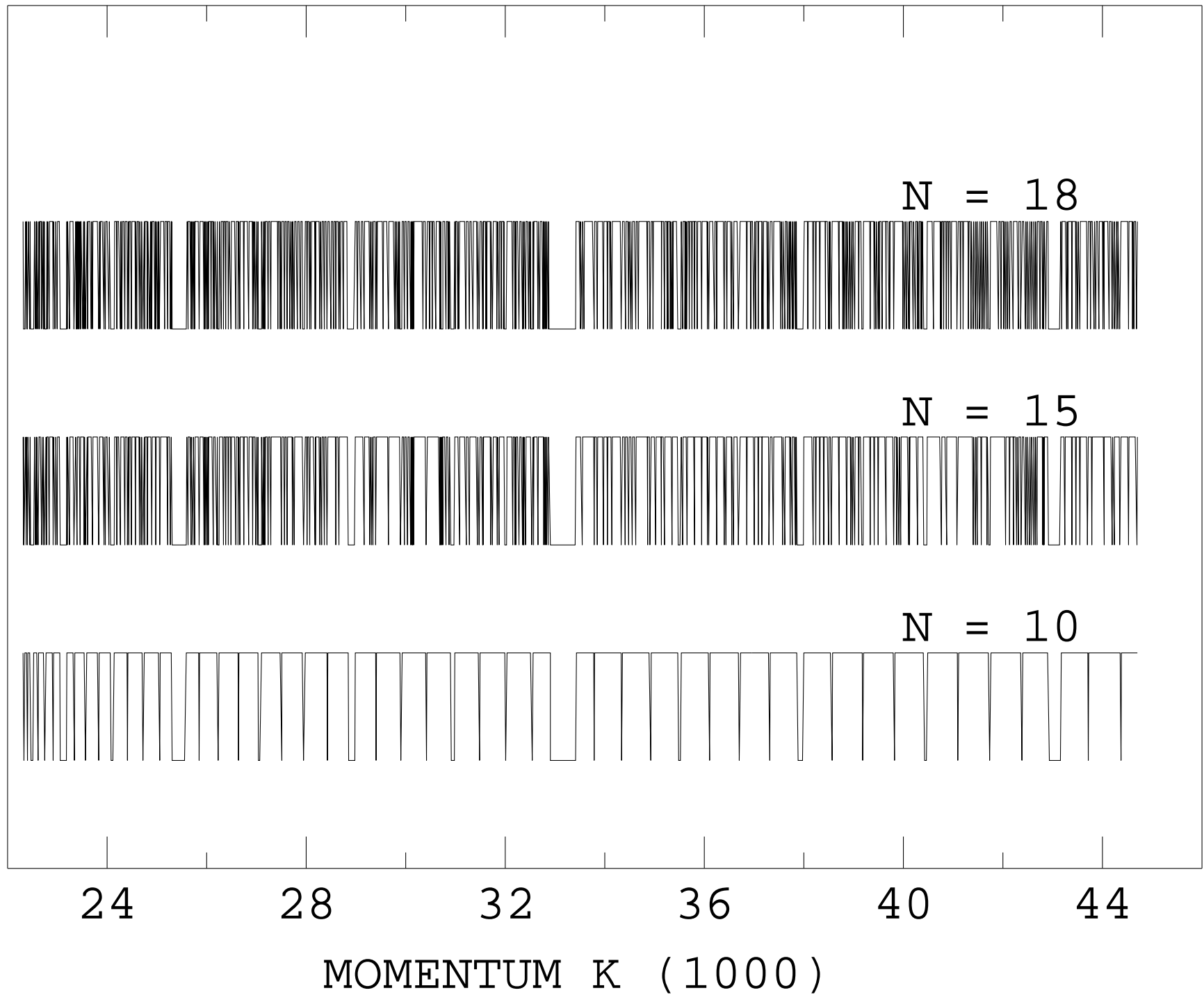
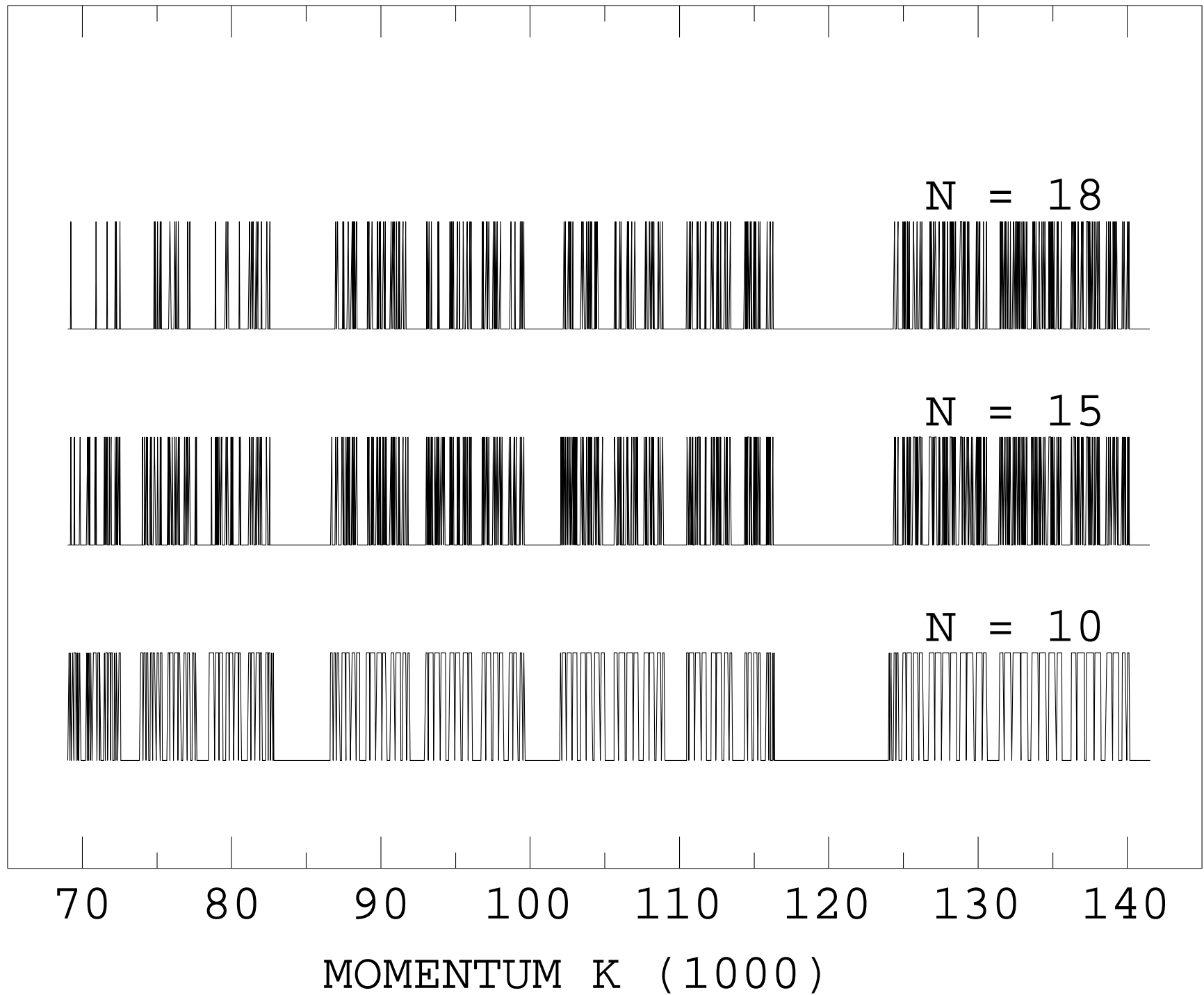
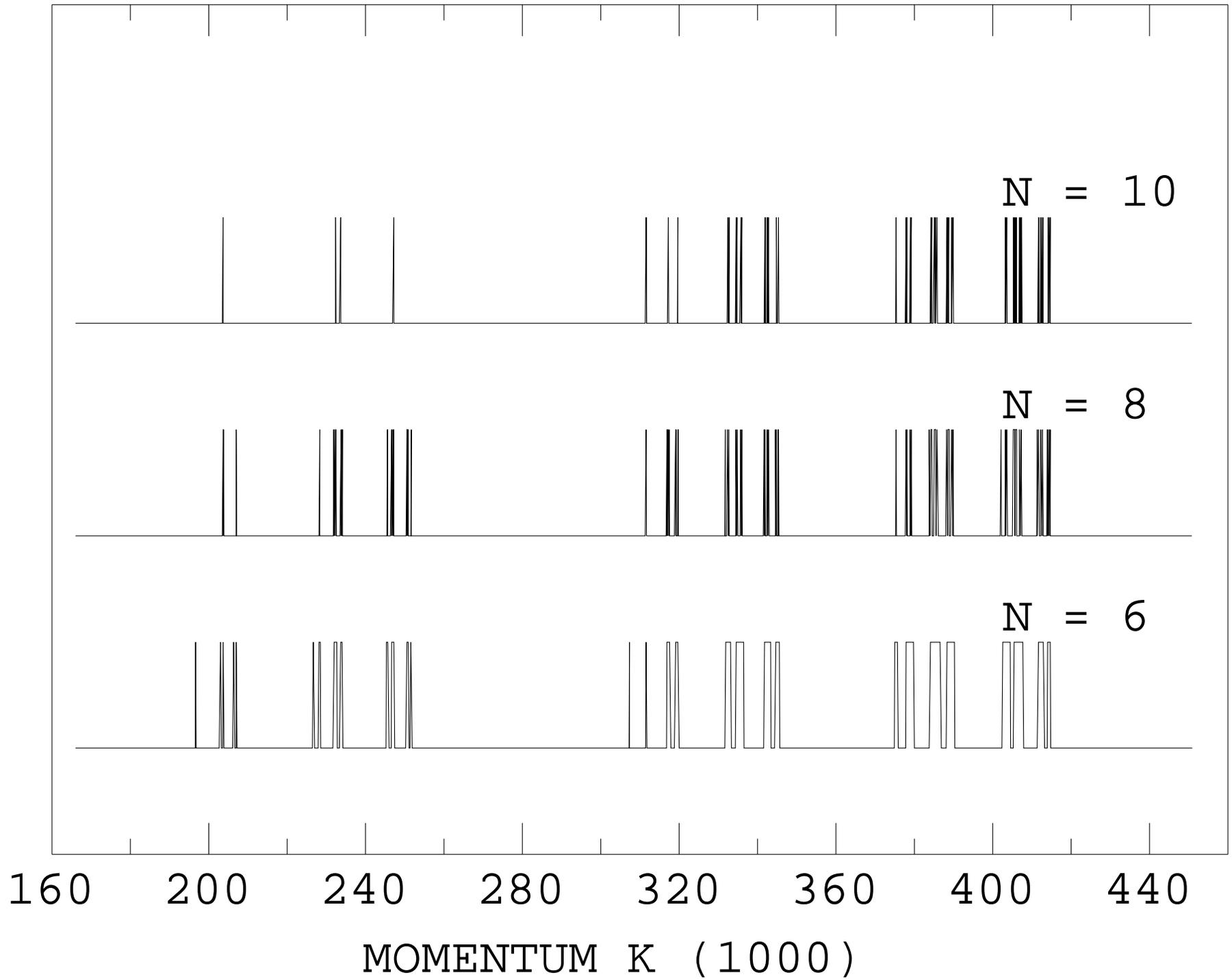
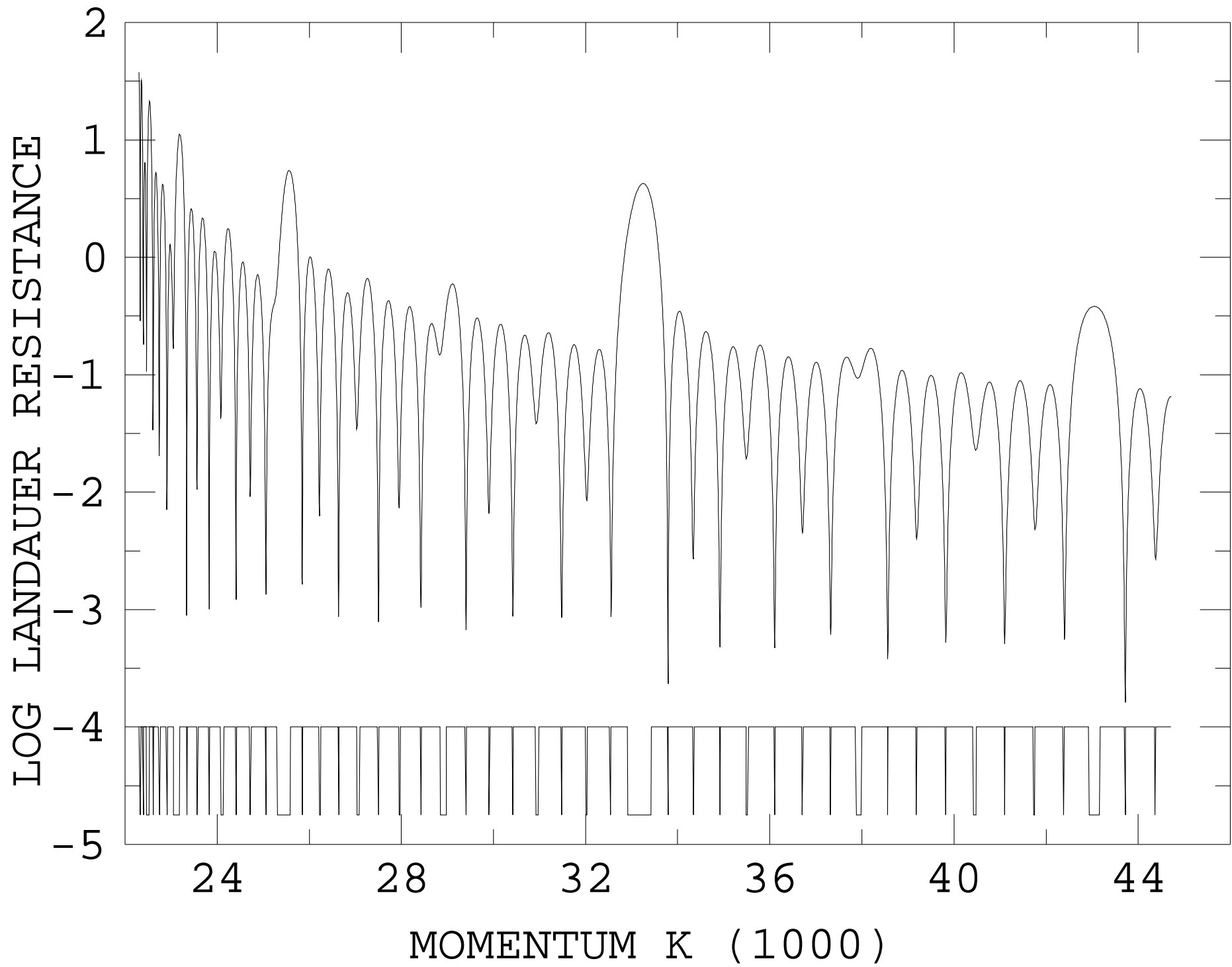
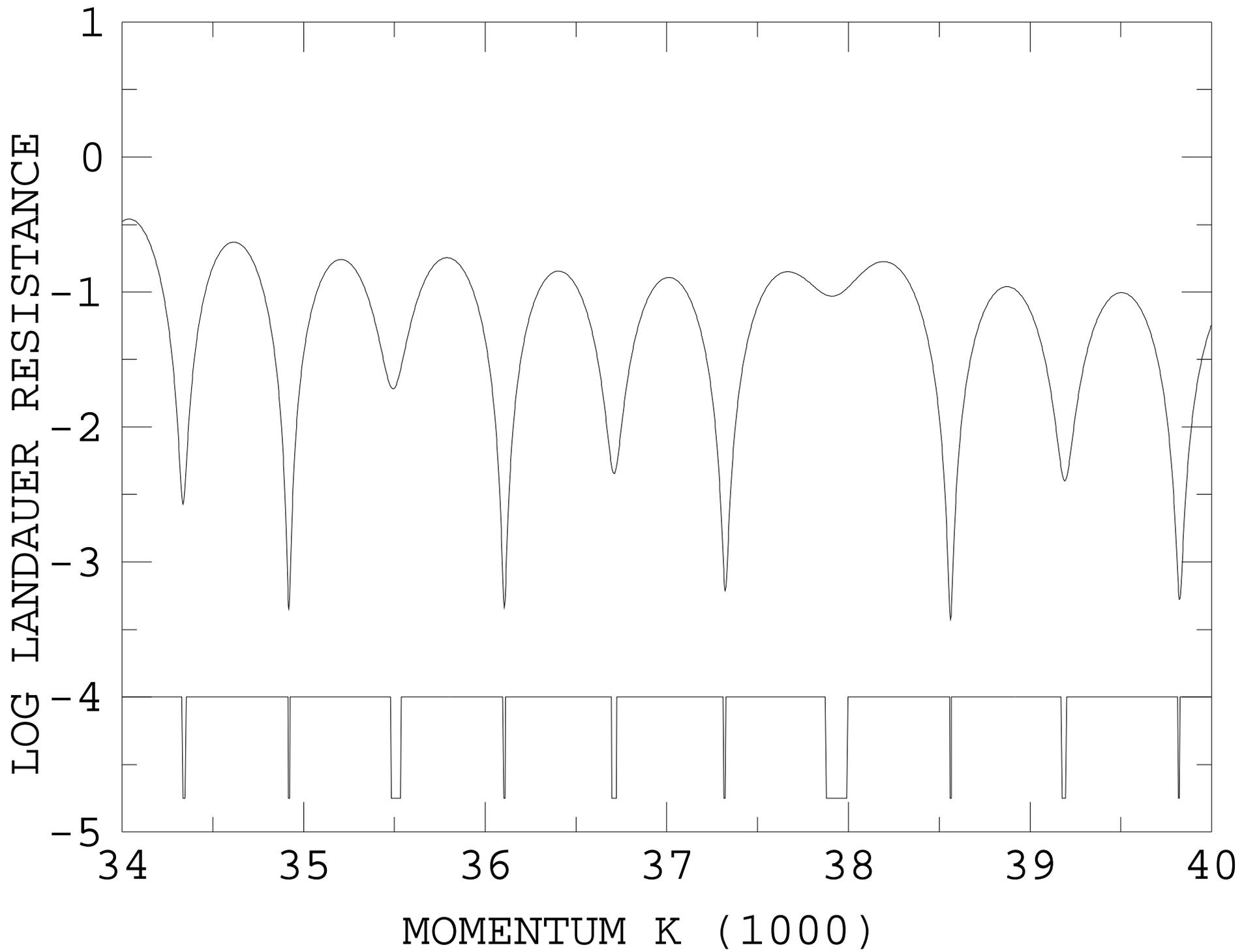
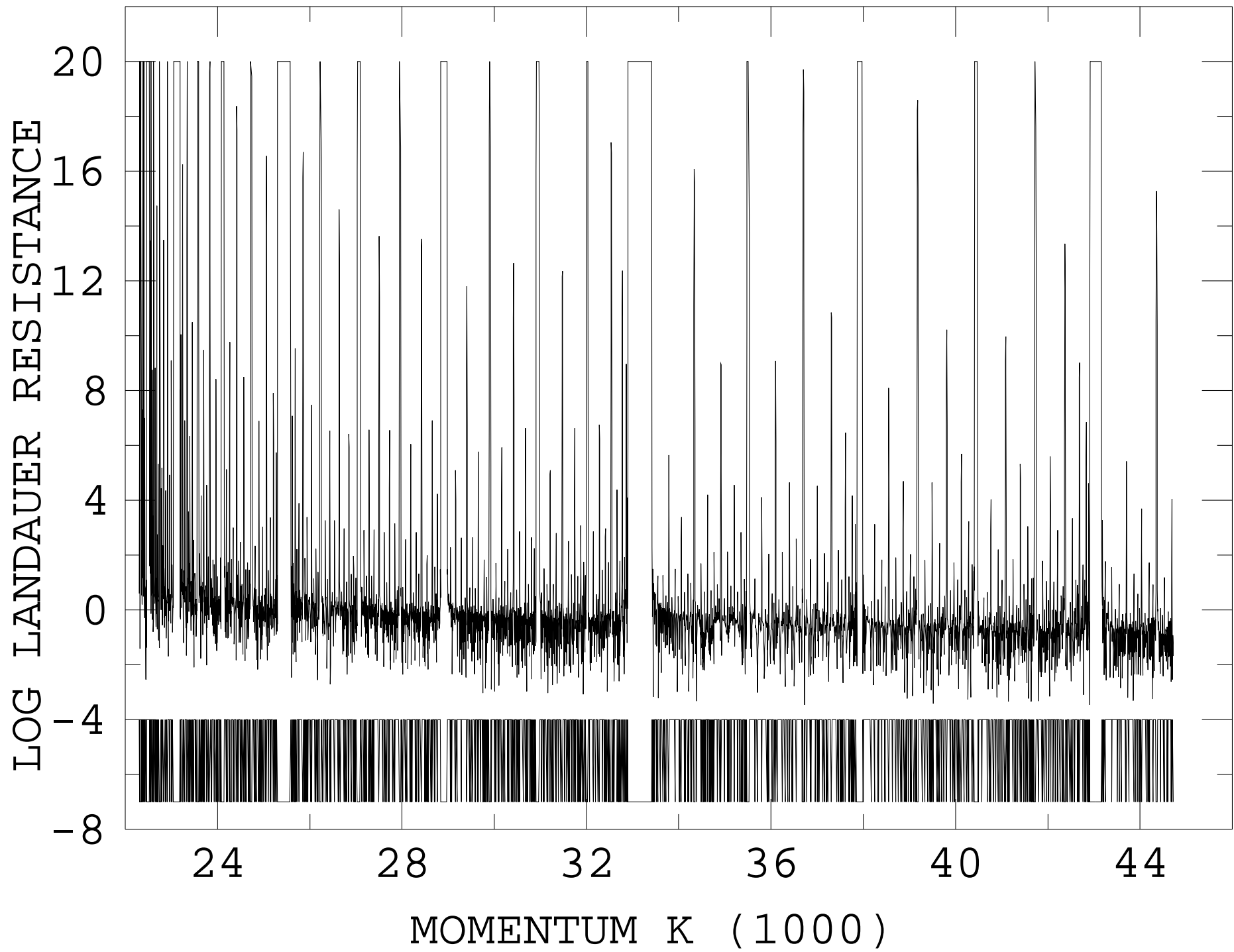
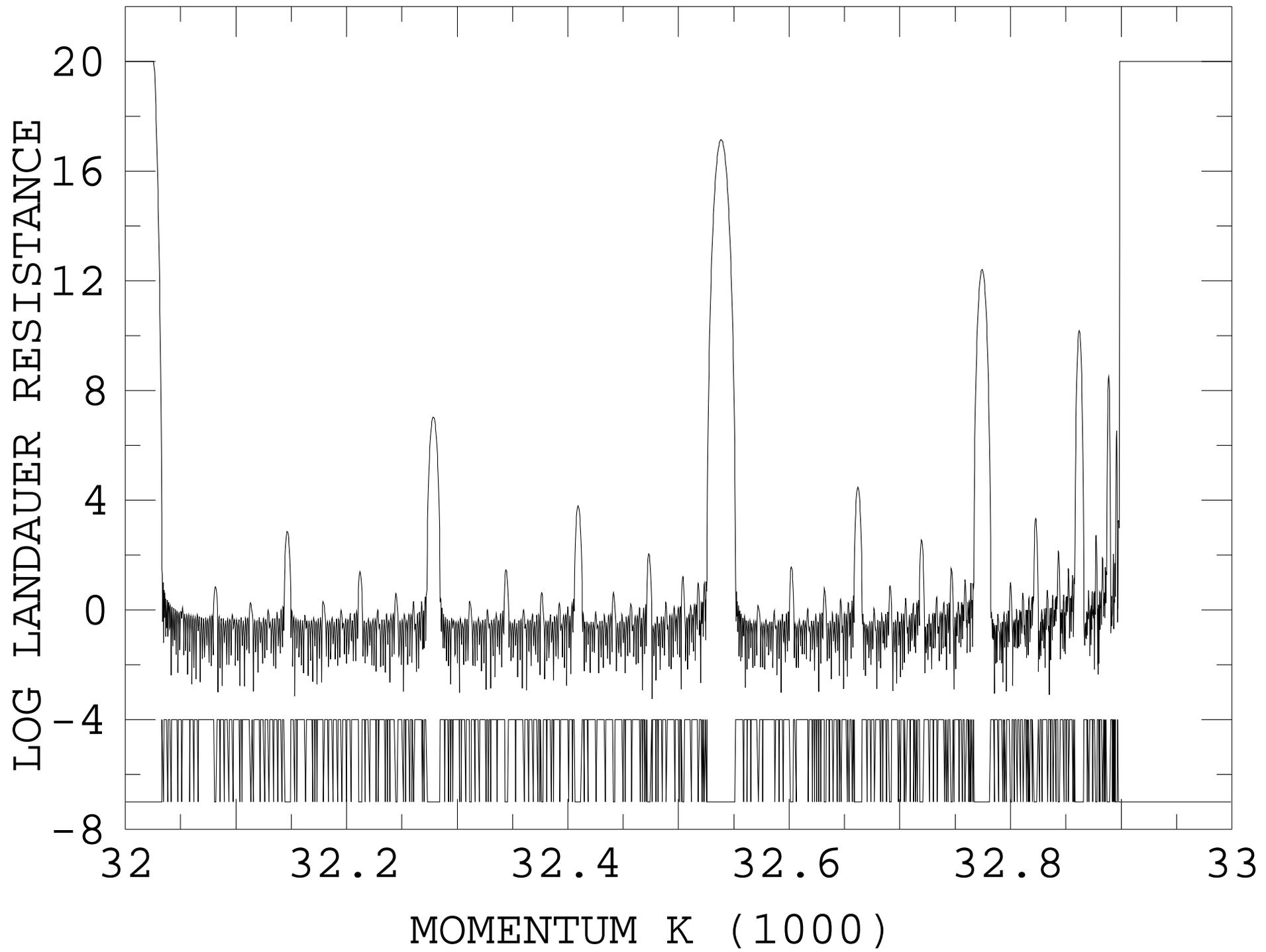
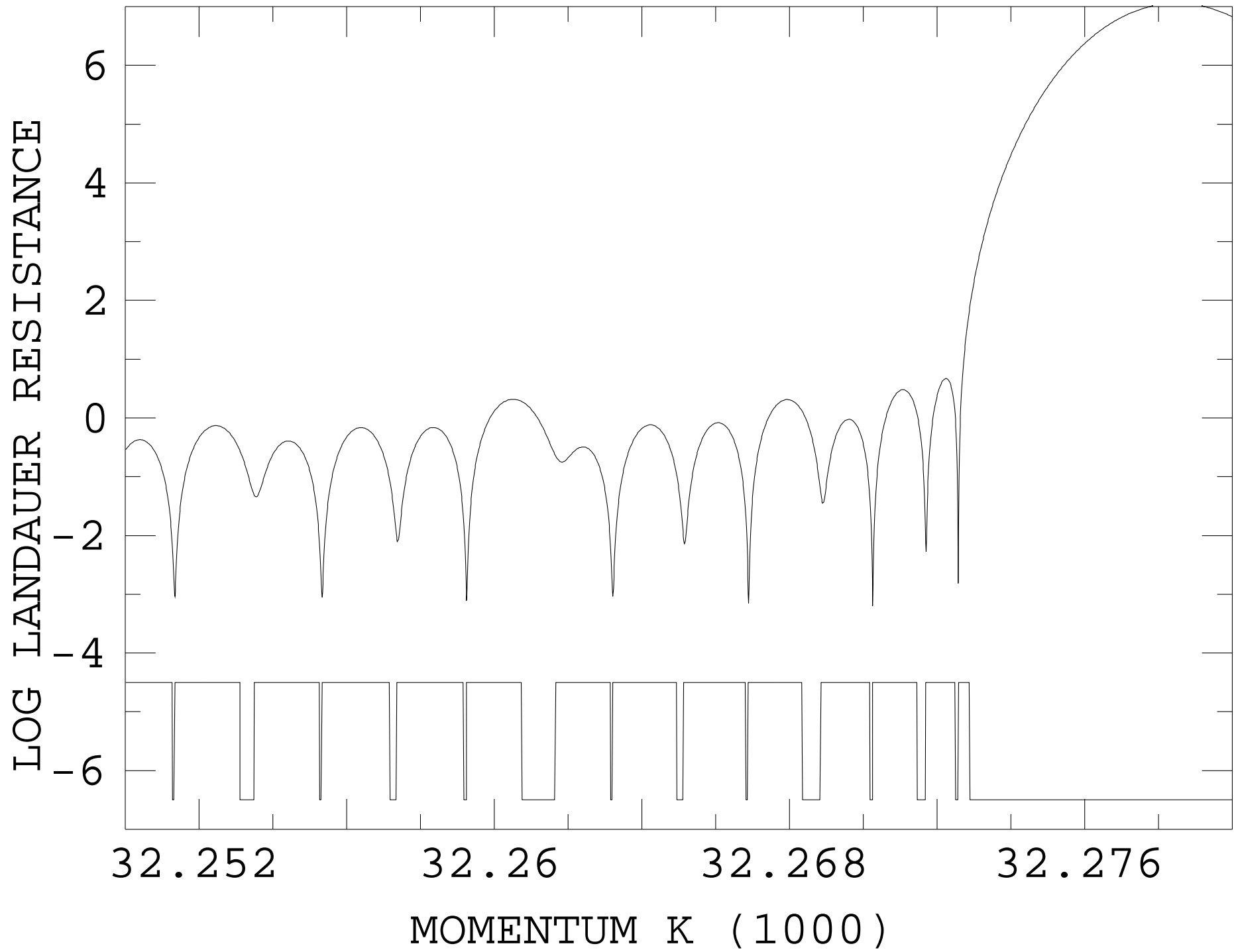
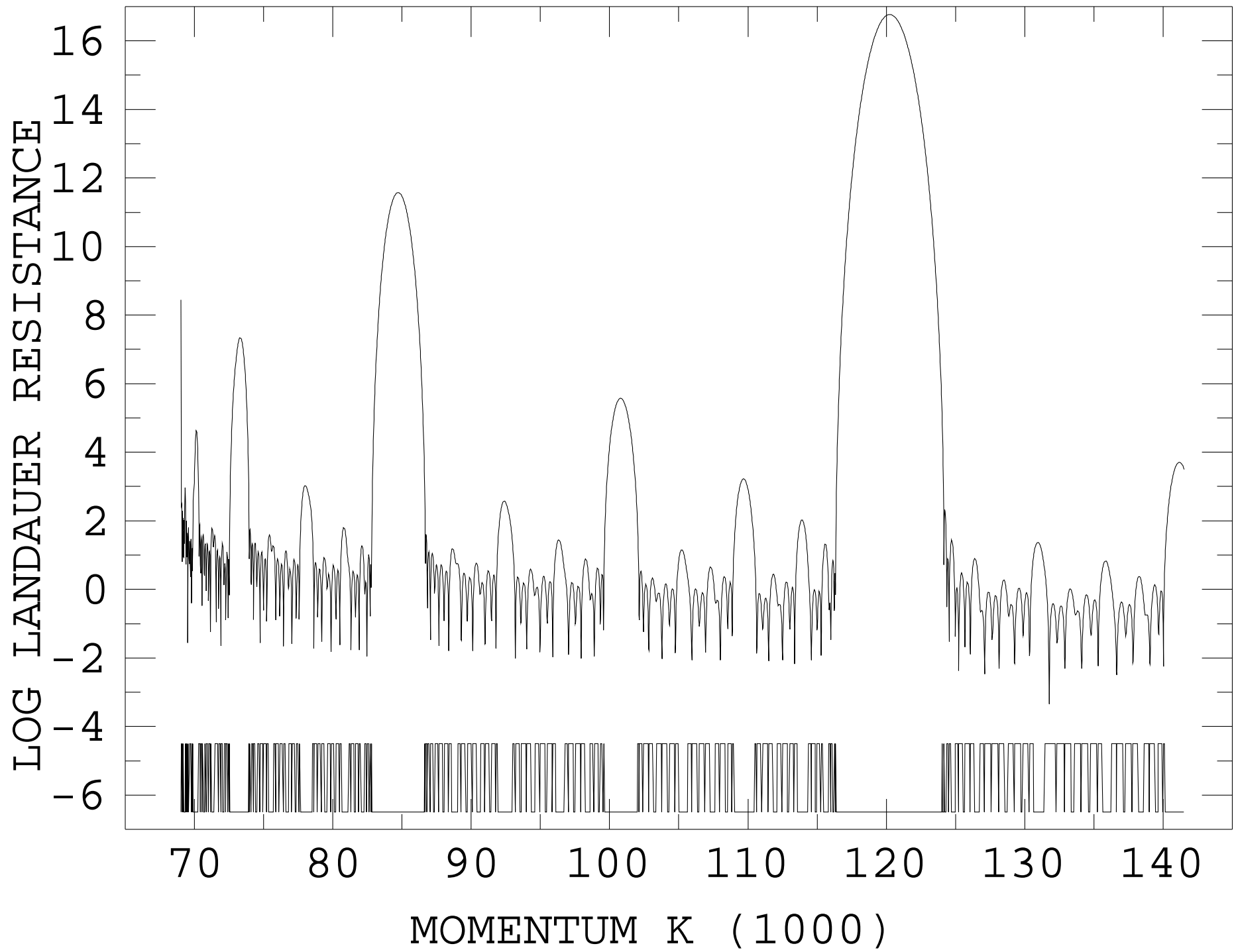
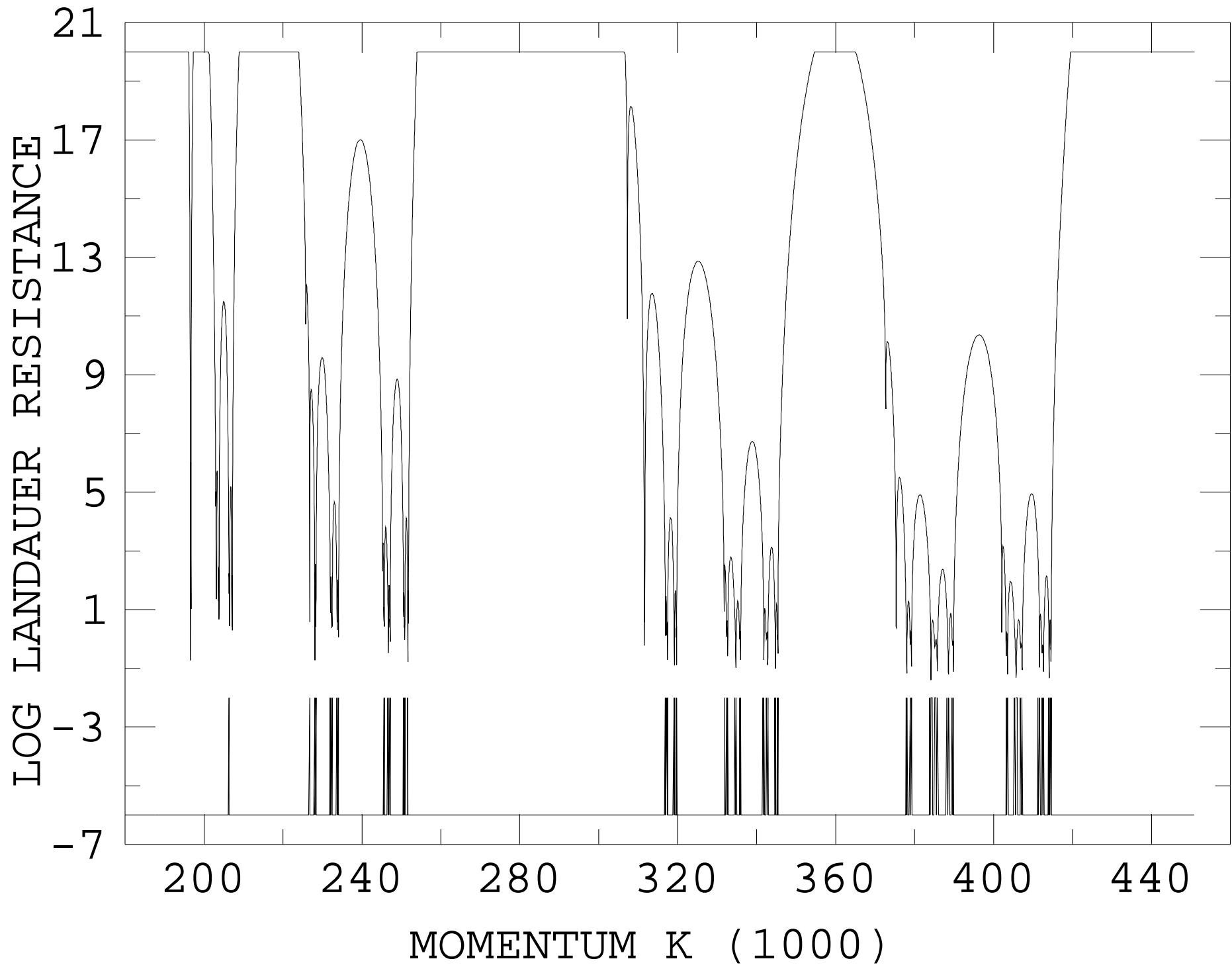
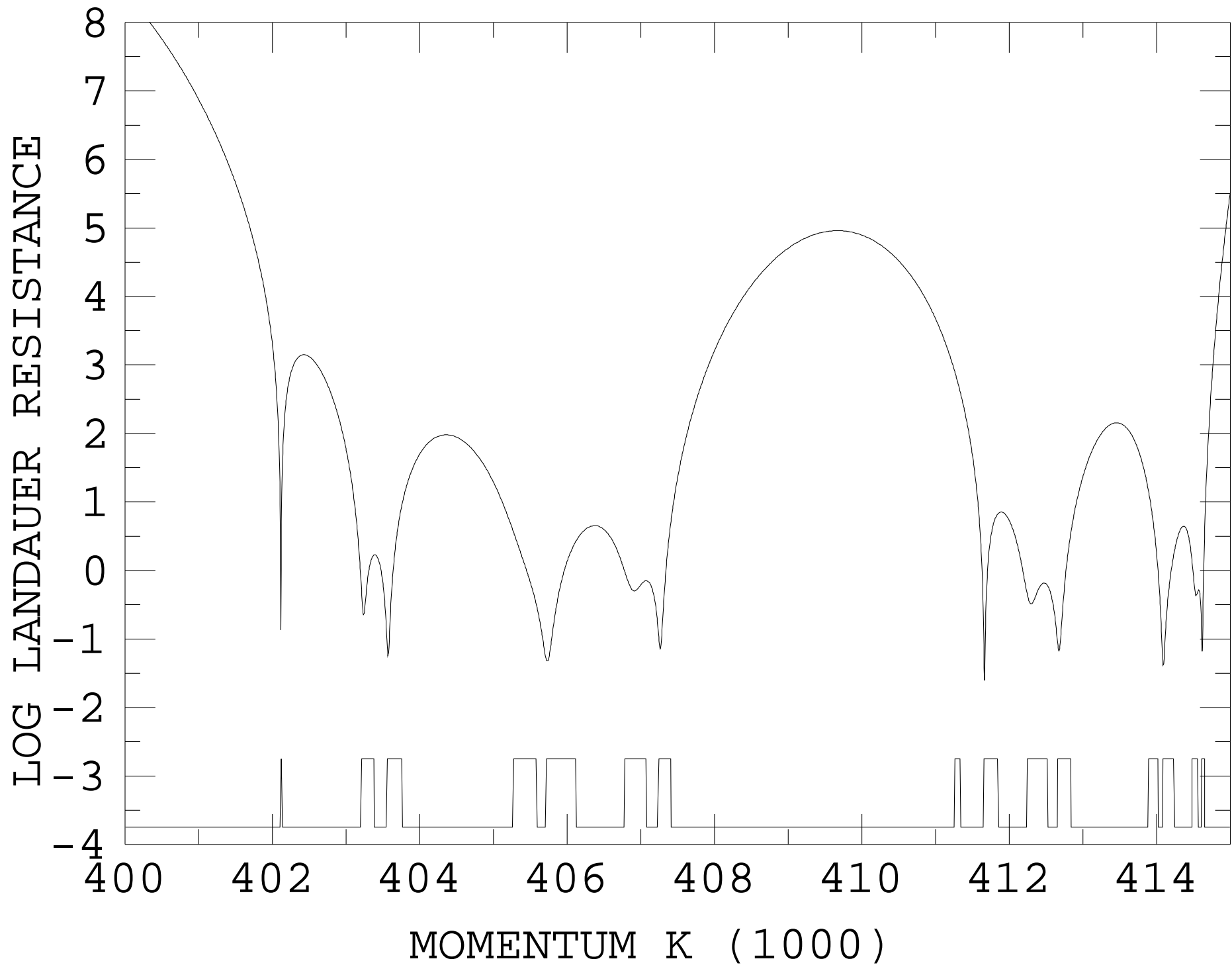
\end{document}